\documentclass[lettersize,journal]{IEEEtran}
\usepackage{amsmath,amsfonts,amssymb}
\usepackage{array}
\usepackage{textcomp}
\usepackage{stfloats}
\usepackage{url}
\usepackage{verbatim}
\usepackage{graphicx}
\usepackage{cite}
\usepackage{textcomp}
\usepackage{xcolor}
\usepackage[noend]{algpseudocode}
\usepackage{subfigure}
\usepackage{amsthm}

\usepackage{hyperref}
\usepackage{comment}
\usepackage{array}
\usepackage{bm}
\usepackage{bbding}
\usepackage{booktabs}
\usepackage{makecell}
\usepackage[frozencache,cachedir=.]{minted}
\usepackage{xcolor}
\usepackage{color}
\usepackage[linesnumbered,lined,commentsnumbered,ruled]{algorithm2e}
\def\BibTeX{{\rm B\kern-.05em{\sc i\kern-.025em b}\kern-.08em
    T\kern-.1667em\lower.7ex\hbox{E}\kern-.125emX}}

\SetAlFnt{\footnotesize}

\newcommand{\tcr}{\textcolor{black}}

\newcommand{\system}{MemFerry\xspace}
\newcommand{\systemMem}{GO-MemFerry\xspace}

 \SetKwInput{KwInput}{Input}              
 \SetKwInput{KwOutput}{Output} 
 \SetKwProg{Fn}{Function}{:}{}

\begin{document}

\title{Fast and Memory Efficient Offload Training Framework with Hybrid XPU Computation\\

% \author{Zhiyi Yao, Zuning Liang, Yuedong Xu, Jin Zhao, Jessie Hui Wang, Xiaowen Chu}
% ? using ?
}

% \and
% \and
% \and
% \and
% \and

\author{\IEEEauthorblockN{Zhiyi Yao, Zuning Liang, Yuedong Xu, Jin Zhao, Jessie Hui Wang, Tong Li}
% Email: $^*$$\{$zyyao23, 21307140019$\}$@m.fudan.edu.cn, $^{*\S}$$\{$ydxu, jzhao$\}$@fudan.edu.cn,\\ $^\dagger$$\{$jessiewang$\}$@tsinghua.edu.cn, $^\ddagger$$\{$tong.li$\}$@ruc.edu.cn
% Email: $^*$$\{$zyyao23, 21307140019$\}$@m.fudan.edu.cn, $^{*\S}$$\{$ydxu, jzhao$\}$@fudan.edu.cn,\\ $^\dagger$$\{$jessiewang$\}$@tsinghua.edu.cn, $^\ddagger$$\{$tong.li$\}$@ruc.edu.cn

\thanks{Manuscript created November, 2026. Zhiyi Yao and Zuning Liang are with College of Future Information Technology, Fudan University; Jessie Hui Wang is with Institute for Network Sciences and Cyberspace, Tsinghua University; Tong Li is with School of Information, Renmin University of China; Yuedong Xu and Jin Zhao are with College of Computer Science and Artificial Intelligence, Fudan University (e-mail: ydxu@fudan.edu.cn).}
}

\maketitle

\begin{abstract}
With the ever-growing size of deep learning models, GPU memory is prone to being insufficient during training. A prominent approach is ZeRO-Offload, which moves the optimizer states to CPU memory and performs parameter update using CPU. However, the deficiencies of ZeRO-Offload include low GPU utilization, imperfect overlapping of communication and computation, and inflexible offloading. In this paper, we leverage Direct Host Access (DHA) on the GPU that can compute data in CPU memory, forming a novel hybrid on-GPU and DHA. We design and implement \system consisting of an execution scheduler and a shadow model. The scheduler strategically chooses layers of parameters for DHA computation and transmits the remaining parameters to GPU memory simultaneously to shorten forward propagation time, and further loads DHA parameters to GPU memory to reduce backward propagation time. The shadow model presents a unified memory abstraction for the parameter partitions stored separately in GPU and CPU memories. To further reduce GPU memory usage, we present \systemMem along with its dynamic programming algorithm that offloads gradients to CPU memory via DHA. We further extend MemFerry to emerging scale-up domains with ScaleUp-MemFerry, which exploits otherwise underutilized accelerator interconnect bandwidth to assist host-to-accelerator data movement through adaptive multi-path transfer. Our experiments show that \system trains up to $1.68\times$ faster and \systemMem can train $1.52\times$ larger model compared to ZeRO-Offload on a single GPU, and increase training speed by at least $28.1\%$ when scaling to data parallelism on 8 GPUs. We further extend the design to a Huawei CloudMatrix384 scale-Up node with up to 8 NPUs, and our ScaleUp-MemFerry reduces the end-to-end iteration time by up to 20.7\% over DeepSpeed. 
%On a CloudMatrix384 scale-up node, ScaleUp-MemFerry further reduces the end-to-end iteration time by up to 20.7\% over DeepSpeed.
\end{abstract}


% \begin{IEEEkeywords}
% component, formatting, style, styling, insert
% \end{IEEEkeywords}

\section{Introduction}
\label{sec: introduction}

% Background of large model training and emphasis on the challenge of memory constraints

With the penetration of large-scale deep learning such as ChatGPT (GPT-3.5 and GPT-4) \cite{DBLP:journals/corr/gpt4} in our daily life and production activities, how to train deep learning models efficiently has attracted great interest. In recent years, the size of deep learning models has increased rapidly. For instance, BERT-base \cite{DBLP:conf/nips/bert} developed in 2018 contains 110 million parameters, while one year after GPT-2 \cite{gpt2} contains 1.5 Billion parameters, and today’s LLaMA \cite{DBLP:journals/corr/LLama} and GPT-3 \cite{DBLP:conf/nips/gpt3} contain 65 Billion and 175 Billion parameters, respectively. However, GPU as the mainstream computing accelerator, have very limited device memory. The latest Nvidia A100, H100 and H800 have 80$G$ Bytes of high-bandwidth memory (HBM), and the frequently used V100 has only 32$G$ Bytes of memory. This imposes a significant challenge on large-scale model training. 

To reduce GPU memory requirements, two key technical routines have surfaced. 
One is to segment the parameters, gradients, optimizer states and activations (also called training content) across multiple GPUs, such as ZeRO~\cite{DBLP:conf/sc/zero} for data parallelism (DP) and GPipe~\cite{DBLP:conf/nips/GPipe} for pipeline parallelism. 
The other is offload training~\cite{DBLP:conf/asplos/SwapAdvisor, DBLP:conf/hpca/Sentinel, DBLP:conf/ppopp/Superneurons, DBLP:conf/usenix/zero-offload}. 
The basic idea is to move partial training content from GPU memory to CPU memory~\cite{DBLP:journals/corr/L2L}, which is usually an order of magnitude larger, or even to secondary storage~\cite{DBLP:conf/sc/zero-infinity, DBLP:conf/sc/STRONGHOLD}. 
However, accommodating larger models comes at the cost of reduced training speed. 
The most widely used offload training system is ZeRO-Offload~\cite{DBLP:conf/usenix/zero-offload}. 
In ZeRO-Offload, the optimizer states are placed in CPU memory and parameter updates are performed by the CPU, while regular forward propagation (FP) and backward propagation (BP) are executed on the GPU. 
Therefore, in each iteration, updated parameters are transferred to GPU memory for FP, and the resulting gradients are returned to CPU memory for parameter updates. 
The limited bandwidth between CPU and GPU thus throttles effective GPU utilization. 
Meanwhile, this bottleneck becomes more pronounced in emerging scale-up domains, where accelerator-to-accelerator connectivity is substantially enhanced while host-to-accelerator data movement remains comparatively constrained, leaving part of the high-bandwidth scale-up fabric underutilized during offload training~\cite{NVL72, AWSTrainium, DBLP:conf/hotchips/JouppiL25, DBLP:journals/corr/abs-2506-12708}.

In this paper, we conduct a preliminary evaluation on the DeepSpeed implementation of ZeRO-Offload with three key observations. (1) GPU streaming processors (SMs) are severely under-utilized compared to non-offloading (e.g. 63\% vs 99\% in BERT-base on Nvidia A100). (2) Communication and computation are not well overlapped. The GPU idleness becomes prominent when the parameters are enormous and unbalancedly distributed on different layers. (3) The memory offloading is not flexible, especially that no further training content can be moved to CPU memory. Hence, we raise a critical question: ``Does there exist a new or improved training paradigm that can achieve higher training speed over ZeRO-Offload and reduce GPU memory usage at the same time?’’

%However, our evaluation shows that the above offload training systems still have shortcomings: 1) GPU under-utilization. GPU computation SM has long idles which means GPU computing power has not been fully utilized. 2) The communication time is not fully hidden in computation. Due to the large size of the model itself, communication time accounts for a large proportion of training. Existing systems ignore completely hidden communication, resulting in low training speed. 3) Inflexible memory offload. The research goal of this paper is to propose a new fast, memory-efficient, and flexible offload training paradigm to address the above problems.

%We propose to integrate the Direct-Memory-Access (DHA) functionality provided by mainstream GPUs in offload training. With DHA, GPU SMs access data in CPU memory, bypassing CPU cores and GPU memory. CNN-DHA \cite{DBLP:conf/glvlsi/CNN-DHA} used DHA to train convolutional neural networks for cache efficiency. Min et al. \cite{DBLP:journals/pvldb/GNN-DHA} trained graph convolutional networks with DHA in order to reduce data loading time. To the best of our knowledge, DeepPlan \cite{DBLP:conf/eurosys/DeepPlan} is the first work to introduce DHA into offloaded model \emph{inference}. Inspired by these studies, we evaluate the 

We propose to integrate the Direct-Host-Access (DHA) functionality provided by mainstream GPUs in offload training. With DHA, GPU SMs access data in CPU memory, bypassing CPU cores and GPU memory. DeepPlan \cite{DBLP:conf/eurosys/DeepPlan} is the first work to introduce DHA into model \emph{inference}. Inspired by this, we evaluate the FP and BP times under the on-GPU and DHA execution modes of offload training. In FP, DHA outperforms on-GPU on certain large layers considerably. In BP, DHA is inferior to on-GPUs if the resulting gradients are stored in GPU memory, but their gaps are greatly reduced if the gradient transmission times of on-GPU are considered. Therefore, fusing on-GPU and DHA into a novel hybrid training paradigm has the potential to achieve our objectives.

We design \system, a novel hybrid training framework that exploits the benefits of DHA in FP while mitigating its side effects in BP as much as possible. 
\system consists of an \textit{execution scheduler} and a \textit{shadow model}. 
The execution scheduler leverages DHA to compute selected layers while loading the remaining parameters into GPU memory during FP, thereby overlapping communication with computation. 
During the interval between the completion of parameter loading and the beginning of gradient transmission, \system further loads selected DHA parameters into GPU memory to reduce their BP time. 
By mitigating communication bubbles in both FP and BP, \system reduces the per-iteration training time. 
The\textit{ shadow model} provides a unified memory abstraction for three parameter partitions spanning CPU and GPU memories: DHA parameters that remain in CPU memory, GFGB (GPU-FP-GPU-BP) parameters that are computed in GPU memory, and DFGB (DHA-FP-GPU-BP) parameters that are computed via DHA during FP but transferred to GPU memory for more efficient BP. 
Through dynamic address mapping, the GPU can transparently access these parameter partitions without additional CPU-side data copies.

To achieve flexible GPU memory offloading, we propose \systemMem that uses DHA to write gradients directly to CPU memory, thus further reducing GPU memory usage. To the best of our knowledge, gradient offloading by DHA has not been reported before. With the user-specified gradient reserving factor, we model the optimal gradient offloading as a dynamic programming problem so as to minimize its impact on the training speed. With \systemMem, we are able to train even larger models and balance the tradeoff between fast training speed and low GPU memory usage. We further adapt MemFerry to emerging scale-up domains by exploiting otherwise underutilized accelerator interconnect bandwidth to assist CPU-to-accelerator data movement. 
ScaleUp-MemFerry combines an additional assisted transfer path with an adaptive data dispatcher that partitions data across the direct and assisted paths according to their latency and bandwidth characteristics.

% \tcr{The evaluation results on clusters consisting of Nvidia V100 and A100 show that \system has faster training speed and lower memory usage compared to the SOTA system during offload training.} \system improves the training speed by $1.32\times$ on average and up to $1.68\times$. The GPU utilization of \system and \systemMem is higher than the baseline by up to $1.27\times$. At the same time, \system reduces training memory usage by up to $61.7\%$. \systemMem reduces the memory usage by $20.1\%$ on average and up to $74.4\%$. With \systemMem, we could train $1.52\times$ larger models compared to the baseline system.
% Furthermore, \system is scalable and compatible with distributed training, with at least $1.28\times$ speedup when we integrate it into data parallelism and a up to $20.7\%$ speedup in emerging scale-up domainss.

The evaluation results on clusters equipped with NVIDIA V100 and A100 GPUs show that \system achieves both faster training and lower GPU memory usage than state-of-the-art offload training systems. 
\system improves training speed by 1.32$\times$ on average and up to 1.68$\times$, while increasing GPU utilization by up to 1.27$\times$. 
\system also reduces GPU memory usage by up to 61.7\%. 
With gradient offloading, GO-\system reduces memory usage by 20.1\% on average and up to 74.4\%, enabling the training of models up to 1.52$\times$ larger than those supported by ZeRO-Offload. 
\system is also scalable to distributed training, achieving at least 1.28$\times$ speedup when integrated with data parallelism on eight GPUs. 
Finally, on a CloudMatrix384 scale-up node, ScaleUp-\system further reduces the end-to-end iteration time by up to 20.7\% over DeepSpeed and 4.0\% over the original \system.

% The evaluation results show that \system and \systemMem have faster training speeds and lower memory usage compared to the SOTA system during offload training. \system improves the training speed by $1.32\times$ on average and up to $1.5\times$. The GPU utilization of \system and \systemMem is higher than the baseline by up to $1.27\times$. At the same time, the \system reduces training memory usage by up to $61.7\%$. \systemMem reduces the memory usage by $20.1\%$ on average and up to $74.4\%$. With \systemMem, we could train $1.52\times$ larger models compared to the baseline system.

%The evaluation results show that \system and \systemMem have faster training speeds and lower memory usage compared to the SOTA system during offload training. \system improves the training speed by $1.32\times$ on average and up to $1.5\times$. The GPU utilization of \system and \systemMem is higher than the baseline by up to $1.27\times$. At the same time, the \system reduces training memory usage by up to $61.7\%$. \systemMem reduces the memory usage by $20.1\%$ on average and up to $74.4\%$. With \systemMem, we could train $1.52\times$ larger models compared to the baseline system.

%\newpage
\section{Background}
\label{sec:background}

\subsection{Memory Usage of DNN Training}
\label{sec:background/Cost of Deep Learning}

We first investigate the memory usage of standard DL training on the premise of sufficient GPU High Bandwidth Memory (HBM). \tcr{All the involved data is usually in the form of tensors}. The data that needs to be maintained on GPU memory (interchangeable with GPU HBM) during training includes: 
1) \textbf{Parameters.} Before the training starts, the parameters of a model need to be loaded into GPU memory. This memory usage is exactly the model size. 
2) \textbf{Gradients.} During the backward propagation (BP) stage, the DL framework will generate gradient tensors of the same size as the parameters on the GPU memory. 
3) \textbf{Optimizer States.} Optimization algorithms in DL usually rely on first-order and second-order momentum to accelerate convergence. \tcr{For example, the optimizer states of Adam \cite{DBLP:journals/corr/Adam} include momentum and variance, each of which is two times the model size.}
4) \textbf{Activations.} \tcr{The activations include the intermediate computing results of each model layer and the loss during BP. If the model input is complex, especially in large language models, activations will occupy a large portion of GPU memory. Fortunately, the recomputation \cite{DBLP:journals/corr/recompute} technique discards intermediate activations when the computation completes so that only the peak size of activations is counted.}

\subsection{ZeRO-Offload Training Paradigm}
\label{sec:background/ZeRO-Offload Training Paradigm}

\begin{figure}%[htbp]
	\includegraphics[scale=0.35, height = 3cm]{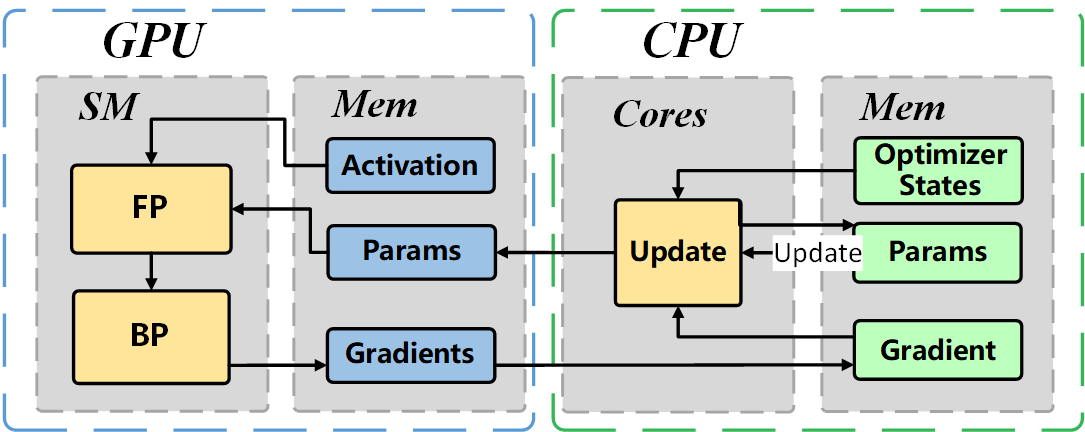}
	\centering
    \vspace{-0.3cm}
	\caption{Training process of ZeRO-Offload.}
	\label{fig:background/zero-offload}
 \vspace{-0.5cm}
\end{figure}

To reduce GPU memory requirement, ZeRO-Offload \cite{DBLP:conf/usenix/zero-offload} presents a new training framework that uses CPU memory to supplement the limited GPU memory. The core idea of ZeRO-offload is to offload the expensive optimizer states and corresponding update computation to CPU. ZeRO-Offload is officially implemented in DeepSpeed \cite{DBLP:conf/kdd/DeepSpeed}, and its training process is shown in Figure \ref{fig:background/zero-offload}.

Before the training starts, the training system stores model parameters and optimizer states in CPU memory, and allocates space for computation in GPU memory. At the beginning of each iteration, CPU explicitly copies the latest model parameters from CPU memory to GPU memory and then performs forward and backward computations on GPU. \tcr{During the BP process, the gradients will be asynchronously transferred to the CPU memory for model update. Once the BP process is finished, the optimizer updates the parameters in the CPU. In this way, the GPU memory consumption of model training is reduced to the size of activation, parameters, and gradients.}

The processing of parameters on the CPU is inefficient because the CPU sequential execution mode is often adopted in deep learning frameworks. The model parameters are physically stored at different parts of the CPU memory such that the parallel processing of them is hard. In view of this, ZeRO-Offload merges all model parameters into a continuous "long" parameter tensor that is stored in a contiguous space. Then, ZeRO-Offload 
develops the CPU-Adam optimization technique by leveraging SIMD on the aggregated tensor-vector. The parallel computing capability of CPU enables ZeRO-Offload to compute more efficiently in the model update. 

\subsection{Scale-up networks}
Scale-up networks have recently emerged as an important architectural trend for large-scale AI systems. Traditional GPU clusters are organized around a scale-out design, where a small group of accelerators within one server are connected by high-bandwidth local links, while communication across servers relies on commodity RNICs and multi-tier Ethernet or InfiniBand fabrics~\cite{DBLP:conf/sigcomm/QianXCGXGFSZMWW24, DBLP:conf/nsdi/JiangLZHCZPLXNJ24}. This hierarchy works well when cross-node traffic is relatively modest, but it becomes increasingly inefficient for modern foundation-model workloads. Once data transfers cross the server boundary, they are exposed to lower bandwidth, higher latency, and CPU-mediated control paths, which limit both scalability and hardware utilization. Scale-up networks address this limitation by extending high-bandwidth accelerator interconnects beyond a single server and building a larger tightly-coupled domain at rack or supernode scale. Within such a domain, tens to hundreds of accelerators can communicate through direct accelerator-centric paths rather than through the conventional host-centric hierarchy~\cite{NVL72, AWSTrainium, DBLP:conf/hotchips/JouppiL25, DBLP:journals/corr/abs-2506-12708}. This architectural shift provides two key benefits. First, it reduces the bandwidth gap between intra-node and inter-node communication, allowing communication-intensive parallelism to scale beyond the physical boundary of one server. Second, it turns the surrounding CPU memory and pooled memory resources into higher-performance memory tiers that can be accessed through the scale-up fabric. As a result, scale-up domains create new opportunities for training systems that must coordinate computation, parameter movement, gradient movement, and optimizer-state placement across heterogeneous memory resources.

\textbf{CloudMatrix384.} A representative production-scale realization of this scale-up design philosophy is Huawei CloudMatrix384~\cite{DBLP:journals/corr/abs-2506-12708, xiao2026huaweicloudmodelasaservicecloudmatrix384}. It integrates 384 Ascend 910 NPUs and 192 Kunpeng CPUs into a unified supernode connected by an ultra-high-bandwidth, low-latency Unified Bus network. Unlike conventional hierarchical clusters, where accelerators are tightly bound to local host CPUs and remote resources are reached through a slower scale-out network, CloudMatrix384 provides a fully peer-to-peer hardware architecture in which NPUs, CPUs, and memory resources can be dynamically pooled and uniformly accessed through the UB fabric. This design is originally motivated by the communication and memory demands of large-scale LLM serving, especially large-scale expert parallelism and distributed KV cache access, but the same hardware substrate is also attractive for memory-efficient training. In particular, CloudMatrix384 exposes a richer set of direct-access paths between accelerators and host-side memory resources. These paths can reduce the cost of moving training states between accelerator memory and CPU memory, and more importantly, they allow the system software to treat host-resident data as an active part of the execution substrate rather than a passive offload target. 

\section{Motivation}
\label{sec: motivation}

In this section, we first present key observations on GPU offloaded training and \tcr{Direct-Host-Access (DHA)}. Then we state our research problem and technical challenges.

\subsection{Limitations of Offload Training}
\label{sec: motivation/Limitations of Offload Training}

We evaluate the performance of GPU offload training in \tcr{DeepSpeed, a widely-used distributed DL framework \cite{DBLP:conf/nips/Pytorch}, and observe three key limitations that motivate our study}.

\noindent\textbf{Low GPU utilization.} In ZeRO-Offload, GPU does not perform computation at every moment. Instead, it resides in the \emph{idle} state for a period of time at each iteration. Take the training of GPT-2 as an example, We find that the idle state occupies around one-third of the GPU runtime, confining the average GPU utilization to be merely $63.1\%$. \tcr{In contrast, when the training is carried out in GPU completely, the GPU utilization is always higher than $80\%$ in every iteration.
The huge utilization gap comes from the data shipment of ZeRO-Offload, where the GPU utilization is near zero.
This indicates that exchanging data between GPU and CPU is quite slow, leading to the waste of expensive GPU resources.} 

\begin{figure}%[htbp]
	\includegraphics[scale=0.3]{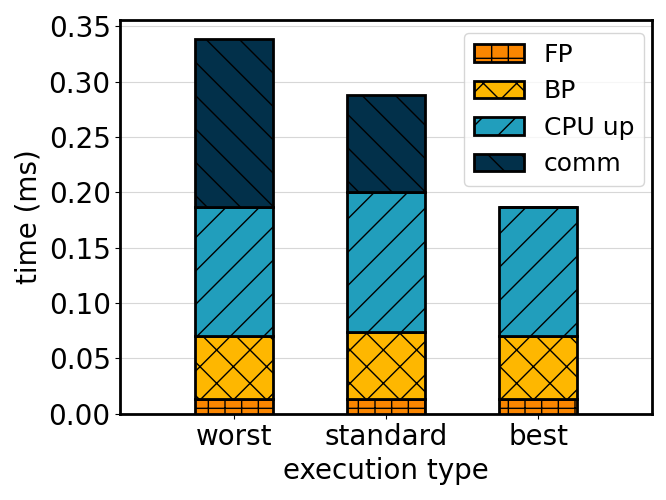}
	\centering
 \vspace{-0.3cm}
	\caption{ZeRO-Offload in an iteration: worst, standard, best.}
	\label{fig:critical path of offload}
 \vspace{-0.5cm}
\end{figure}

{\color{black}
\noindent\textbf{Insufficient communication overlapping.} We delve into the execution schedule of ZeRO-Offload to understand the root cause of GPU under-utilization. ZeRO-Offload leverages the overlapping of communication and computation to reduce the offloading overhead. \tcr{As shown in Figure \ref{fig:motivation/ZeRO-Offload's execution schedule}, the backward propagation (labeled as `BP’) at GPU partially overlaps the gradient synchronization from GPU to CPU (labeled as `Gradients'), and the parameter update (labeled as `Update') partially overlaps the parameter synchronization from CPU to GPU (labeled as `Params'). However, such overlapping is imperfect with two bubbles observed in each iteration. Owing to the layer-wise execution of FP and BP, the communication of a layer always takes place after the completion of its computation, generating bubbles in the figure.} Even worse, during BP, the final computed layer is often the embedding layer in large language models, which have large model sizes and long transmission latency.

Fig.\ref{fig:critical path of offload} illustrates the drawback of ZeRO-Offload execution schedule and highlights how good the optimal overlapping can be. ZeRO-Offload is evaluated under three scenarios: “worst” means that there does not exist the overlapping of communication and computation; “standard” means the normal operation; “best” is the direct removal of the communication time from “worst”. Due to the interference between communication and computation, the BP and CPU update times are prolonged in “standard”, yet the communication time is partially hidden in the computation time. But compared to “best” execution, there is still 57.2\% unhidden computation time that leads to training efficiency degradation.

\begin{figure}
\centering
  \subfigure[ZeRO-Offload's execution schedule]{
  \label{fig:motivation/ZeRO-Offload's execution schedule}
  \begin{minipage}{.5\textwidth}
  \centering
    \includegraphics[scale=0.27]{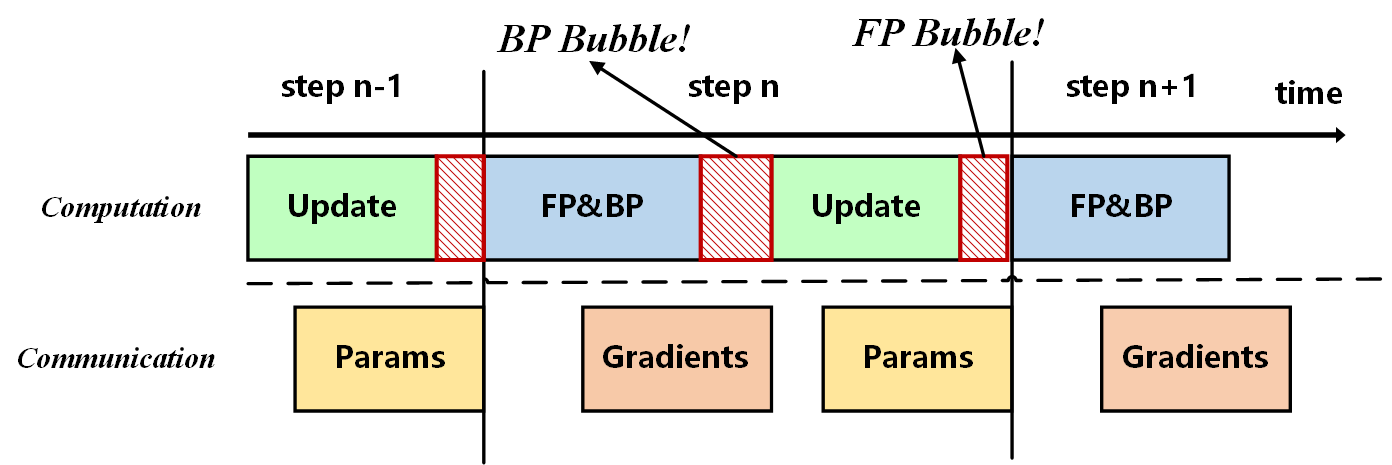}
  \end{minipage}
  }\\
  \vspace{-0.2cm}
  \subfigure[\system's execution schedule]{
  \label{fig:motivation/execution schedule}
  \begin{minipage}{.5\textwidth}
  \centering
    \includegraphics[scale=0.27]{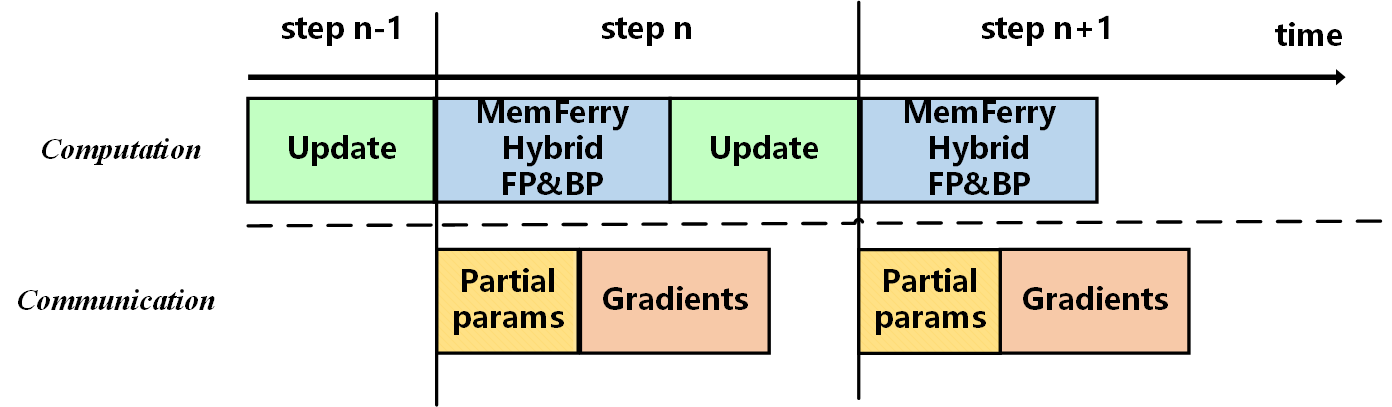}
  \end{minipage}
  }
    \vspace{-0.4cm}
   \caption{Training schedule: ZeRO-Offload vs MemFerry.}
  \label{fig:motivation/Training resource schedule comparison} 
  \vspace{-0.5cm}
\end{figure}

}

\begin{figure*}[!ht]
\centering
\subfigcapskip=0pt
  \subfigure[GPU access data in CPU memory]{
  \label{fig:observation/GPU access data in main memory}
  \begin{minipage}{.2\textwidth}
  \centering
  \vspace{1pt}
    \includegraphics[scale=0.4]{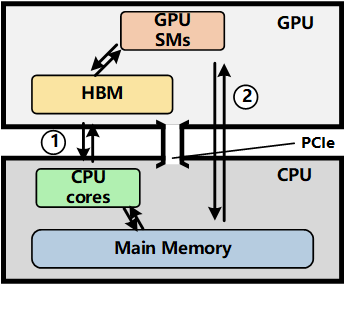}
  \end{minipage}
  }
% \subfigcapskip=0pt
  \subfigure[FP time: on-GPU vs DHA parameters]{
  \label{fig:motivation/FP of DHA parameters}
  \begin{minipage}{.24\textwidth}
  \centering
  % \vspace{12pt}
    \includegraphics[scale=0.23]{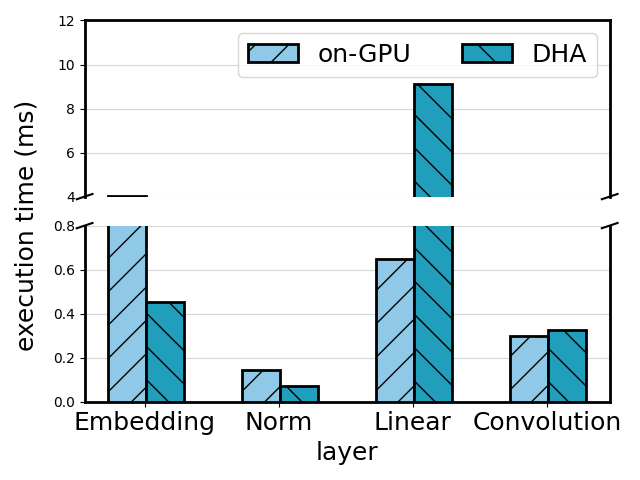}
  \end{minipage}
  }
% \subfigcapskip=0pt
  \subfigure[BP time: in-HBM vs DHA parameters]{
  \label{fig:motivation/BP of DHA parameters}
  \begin{minipage}{.24\textwidth}
  \centering
  % \vspace{12pt}
    \includegraphics[scale=0.22]{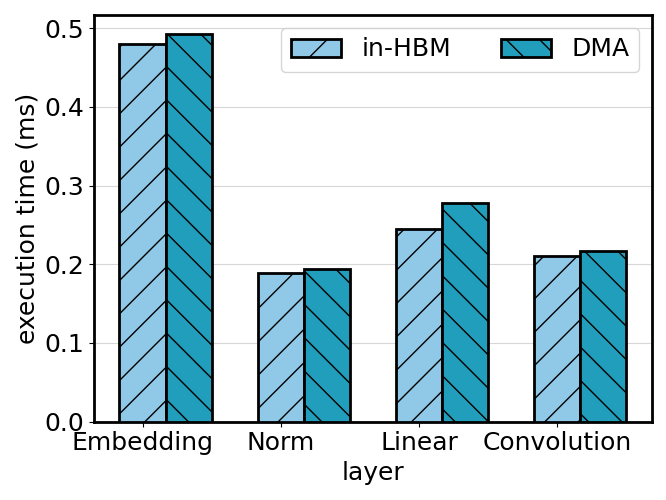}
  \end{minipage}
  }
  \subfigure[BP time: in-HBM vs DHA gradients]{
  \label{fig:motivation/BP of DHA gradient}
  \begin{minipage}{.24\textwidth}
  \centering
  % \vspace{12pt}
    \includegraphics[scale=0.24]{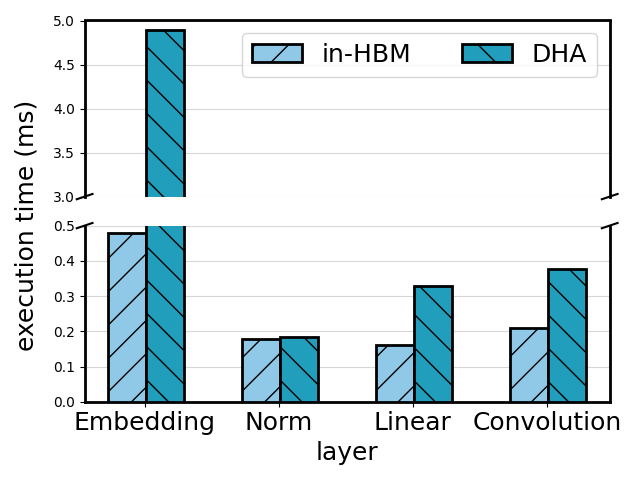}
  \end{minipage}
  }
  \vspace{-0.3cm}
  \caption{DHA and on-GPU execution.}
  \vspace{-0.5cm}
\end{figure*}

{\color{black}
\noindent\textbf{Inflexible GPU offload.} The offloading policy of ZeRO-Offload is inelastic, i.e. a fixed proportion of training contents is moved to CPU memory. \tcr{However, the GPU memory budget needs to be utilized as much as possible to improve the training throughput.} There are contents such as gradients that do not have to be stored in the GPU memory in its entirety, an offload training system could offload them when the GPU memory budget is small. Therefore, memory offloading needs to be flexible, catering to versatile memory sizes.

}

\textbf{Summary}: Despite the reduced memory occupation of ZeRO-Offload, we observe that the GPU utilization is far from the optimality and the memory offloading is not flexible. A more fine-grained memory offloading policy is demanded to improve the training speed and reduce the GPU memory usage in the meantime.

\subsection{GPU Computation on CPU Memory}
\label{sec: motivation/GPU Computation with Host Memory}

\tcr{Commodity GPU streaming multiprocessors (SMs) possess two computing approaches with the data stored in CPU memory \cite{cuda}. 
In Figure \ref{fig:observation/GPU access data in main memory}, we illustrate these two approaches with Nvidia GPU as an example.
\textcircled{1} \textbf{on-GPU}: the data is ferried from CPU memory to GPU memory (i.e. GPU HBM) before it is computed by GPU SMs. 
The advantage of on-GPU is the high bandwidth between GPU memory and SMs, while the cost of CPU memory to GPU memory transmission is non-trivial.
\textcircled{2} \textbf{Direct Host Access (DHA)}: The \textit{CudaHostAlloc} function is called to allocate the space in CPU memory and register it to the GPU. In this way, GPU SMs can directly access the registered CPU memory without
first staging the data in GPU memory. 
Naturally, the computation via DHA is slower compared with on-GPU. However, the DHA computation streams and the CPU-GPU data transfer streams in GPU SM are independent, so they can be executed in parallel.}

\tcr{Existing ML frameworks such as Pytorch \cite{DBLP:conf/nips/Pytorch} do not support DHA-enabled training. Even though ZeRO-Offload partitions the data in both CPU and GPU memories, the FP and BP operations are entirely processed via the on-GPU mode.}
\tcr{DeepPlan \cite{DBLP:conf/eurosys/DeepPlan} introduces DHA computation into model \emph{inference}.} 
With DHA, GPU can load several layers of model parameters from CPU memory to GPU memory, and perform FP computation on other layers in CPU memory in parallel. 
\tcr{To explore the potential of DHA in modeling training, we profile the efficiency of both on-GPU and DHA approaches using \emph{Bert-base} as an example. The results below are measured from a machine with an Nvidia A100 and $8$ CPU cores interconnected via $100Gbps$ PCIe switch. 
We next show the training performance when we offload part of the training contents (e.g. parameters and gradients) to CPU memory and enforce GPU to operate them via DHA.}

\noindent\textbf{DHA on CPU parameters in FP.} We first measure the FP times of on-GPU and DHA execution for the parameters stored in CPU memory in Figure \ref{fig:motivation/FP of DHA parameters}. Note that both the transmission time and the computation time are included in the FP time of on-GPU, while they are inseparable in that of DHA. 
For the \textit{Embedding} and \textit{LayerNorm} layers, DHA exhibits much shorter FP time than on-GPU. \tcr{This is because the transmission time of these layers is longer than the computation time, and the on-GPU execution is blocked by the transmission.} In contrast, the FP times of the \textit{Linear} and \textit{Convolution} layers are longer in DHA than in on-GPU. Obviously, for these layers with high computational loads but small in size, they are more suitable to be executed via the on-GPU approach. 

\noindent\textbf{DHA on CPU parameters in BP.}
If FP uses on-GPU approach (load parameters from CPU memory), then the model parameters are already in GPU HBM during the BP process. Then, GPU is able to directly use the in-HBM parameters to perform BP. If DHA is used to perform FP, the parameters are still in CPU memory which means GPU still needs to use DHA to perform BP. In Figure \ref{fig:motivation/BP of DHA parameters}, we show the BP time of these two approaches. Overall, DHA is slower than on-GPU. While compared to FP, the performance degradation of DHA BP is not severe $(5-10\%)$. This is because deep learning frameworks have activation retention mechanisms. However, completely ignoring the performance degradation of DHA BP can also lead to poor training efficiency. 
% , which results in less computation involving parameters in the BP
% the significant difference in

\noindent\textbf{DHA on CPU gradients in BP.}
\tcr{During the traditional BP process of offload training, gradients will be generated on the GPU HBM and then transmitted to CPU memory.} However, we can make use of DHA to generate gradients directly on CPU memory (leveraging DHA write). So we propose BP integrating DHA gradients as an additional BP execution approach.  Figure \ref{fig:motivation/BP of DHA gradient} shows the measurement of BP time with DHA gradients. Compared to in-HBM BP which directly generates gradients in GPU HBM, BP with DHA gradients executes slower. The reason is that the former's SM can write gradients to HBM with high bandwidth, but the latter needs to write gradients to CPU memory through the low-bandwidth PCIe switch. \tcr{However, DHA gradients do not need any GPU HBM buffer to save gradients, which means the GPU memory requirements for BP are significantly reduced.}

\subsection{An Ideal Execution Schedule}
\label{sec:motivation/exection schedule}

% 1. propose ideal schedule
% 2. DHA solve bubble 1
% 3. gradient transmission parallel with update to solve bubble 2
% 4. transition to next chapter

The above observations manifest that DHA enriches the FP and BP execution approaches so as to possibly reduce the data ferry time between the CPU memory and the GPU memory. 
Here, we elaborate on an ``ideal'' scheduling policy for offloaded training (in Figure \ref{fig:motivation/execution schedule}) and show how to utilize the advantages of the hybrid on-GPU and DHA.

\begin{figure}%[htbp]
	\includegraphics[scale=0.3]{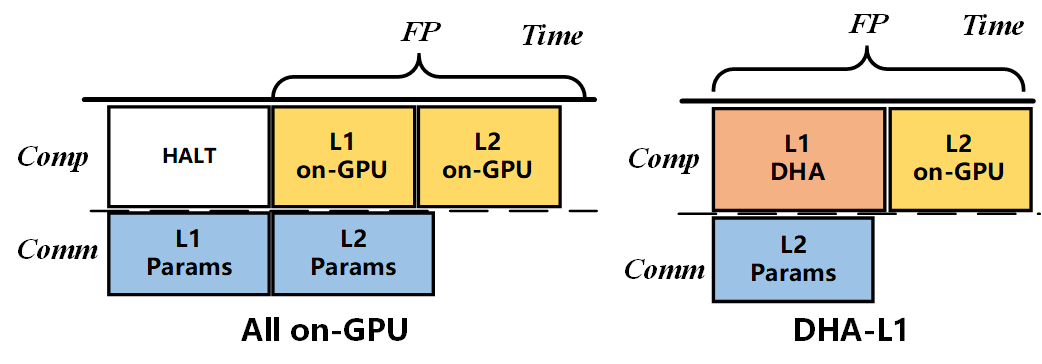}
	\centering
 \vspace{-0.3cm}
	\caption{An example of hybrid execution.}
	\label{fig:motivation/DHA overlap}
 \vspace{-0.5cm}
\end{figure}

\tcr{To minimize GPU idleness, the execution schedule could leverage hybrid DHA and on-GPU computation.} 
Recall in Figure \ref{fig:motivation/ZeRO-Offload's execution schedule}, two bubbles exist in each iteration of ZeRO-Offload. The FP bubble is caused by the layer-by-layer parameter transmission from CPU memory to GPU memory. Hence, we can simultaneously load the parameters of some layers to GPU memory and perform FP computation on other layers in CPU memory via DHA. Figure \ref{fig:motivation/DHA overlap} compares the execution modes of a two-layer model. When the two layers are both performed via on-GPU (labeled as "All on-GPU"), there is a long computation halt of waiting for parameters transmission. But when layer-$1$ FP is conducted via DHA and layer-$2$ is on-GPU, the transmission of parameters is completely hidden and computation starts with no startup latency. In Figure \ref{fig:motivation/execution schedule}, one can see that ``Partial params'' transmission executes in parallel with ``Hybrid FP\&BP''.

The BP bubble (in Figure \ref{fig:motivation/ZeRO-Offload's execution schedule}) appears in the gradient transmission between BP and update. The gradient is always generated after BP even if we use a fine-grained schedule. \tcr{Hence, unlike hybrid execution in FP, we cannot transmit the gradient of a layer before it is generated. However, if gradient transmission could be carefully partitioned, the gradient transmission could perform with parameters update in parallel. This is reflected in Figure \ref{fig:motivation/execution schedule} that the ``gradients" transmission is partly execution parallel with ``BP" and ``update".}

Achieving such an ideal runtime scheduler with flexible GPU memory offloading has three major challenges.

\begin{itemize}
\setlength{\itemsep}{0pt}
\setlength{\parsep}{0pt}
\setlength{\parskip}{0pt}
    \item \textbf{Complex training decisions in hybrid training.} Different layers of DL models have different affinities for the execution approaches. \tcr{We need to design a fine-grained execution schedule leveraging the advantages of hybrid execution training to minimize training time.}
    % GPU in hybrid execution has two execution approaches, DHA and on-GPU computation. 
    
    \item \textbf{Unified management of GPU and CPU memory.} {\color{black} \tcr{Existing frameworks lack efficient storage support to manage normal and  DHA training contents on GPU memory and CPU memory during training.} If we blindly copy data even under an optimal schedule, the additional overhead may offset the performance gain.}
    % Therefore, a unified memory management mechanism is a necessity in hybrid training.
    % Hybrid training introduces GPU-CPU communications back and forth.
    
    % Hybrid training introduces cross-GPU-CPU direct data access. However, existing frameworks do not have perfect runtime hybrid model storage support. If the GPU and CPU models in offload training are roughly separated, it will only bring additional overhead of data copying. We need to provide a unified memory management solution suitable for model training frameworks.
    
    \item \textbf{Flexible GPU memory offloading.} The current offloading training designates fixed contents to be offloaded to CPU memory, \tcr{which is not flexible and makes peak GPU memory usage unchangeable. Further offloading data to CPU memory with a gentle sacrifice of training speed remains a challenge.}

\end{itemize}

\section{System Design}
\label{sec:system design}

% {\color{black}
% In this section, we present an overview of \system with three key \tcr{aspects}, including hybrid training modes, efficient memory management and flexible offloading. }

In this section, we present an overview of MemFerry with four key aspects, including hybrid training modes, efficient memory management, flexible offloading, and scale-up-aware data movement.

\subsection{System Overview}
\label{sec:system design/system design}

We briefly describe our technical measures toward the key challenges raised in Section \ref{sec: motivation}, followed by the workflow.

\noindent\textbf{1) Scheduler for hybrid training.} (Section \ref{sec:system design/training schedule}) \tcr{We design a new offloading execution scheduling with hybrid training that overlaps parameter loading time and gradient transmission time with DHA computation. Our schedule proposes three execution modes}, GFGB, DHA and DFGB that are automatically chosen for different layers. 

\noindent\textbf{2) Runtime shared memory management.} 
(Section \ref{sec:system design/memory management}) We propose ``shadow model'', a logical model on GPU where its underlying storage space spans both GPU memory and CPU DHA memory. \tcr{The shadow model allows all the tensors to seamlessly switch the underlying storage address between DHA memory and GPU memory, avoiding unnecessary data copy at runtime.}

\noindent\textbf{3) On-demand offloading of GPU memory.} 
(Section \ref{sec:system design/gradient offload}) With DHA, not only the parameters but also the gradients can be offloaded to CPU memory. \tcr{The gradient offloading could further reduce GPU memory usage at the cost of the gentle increase in per-iteration training time. This means that an elastic choice in pursuit of fast training speed or low GPU memory usage is allowed.}

\noindent \textbf{ 4)Scale-up-aware data transfer.} (Section \ref{sec:design/scale-up})
We extend MemFerry to emerging scale-up domains by using
accelerator interconnects as assisted paths for CPU-to-accelerator
data movement. An adaptive data dispatcher partitions each
transfer according to the latency and bandwidth characteristics
of the available paths.

 \begin{figure}%[htbp]
	\includegraphics[scale=0.35]{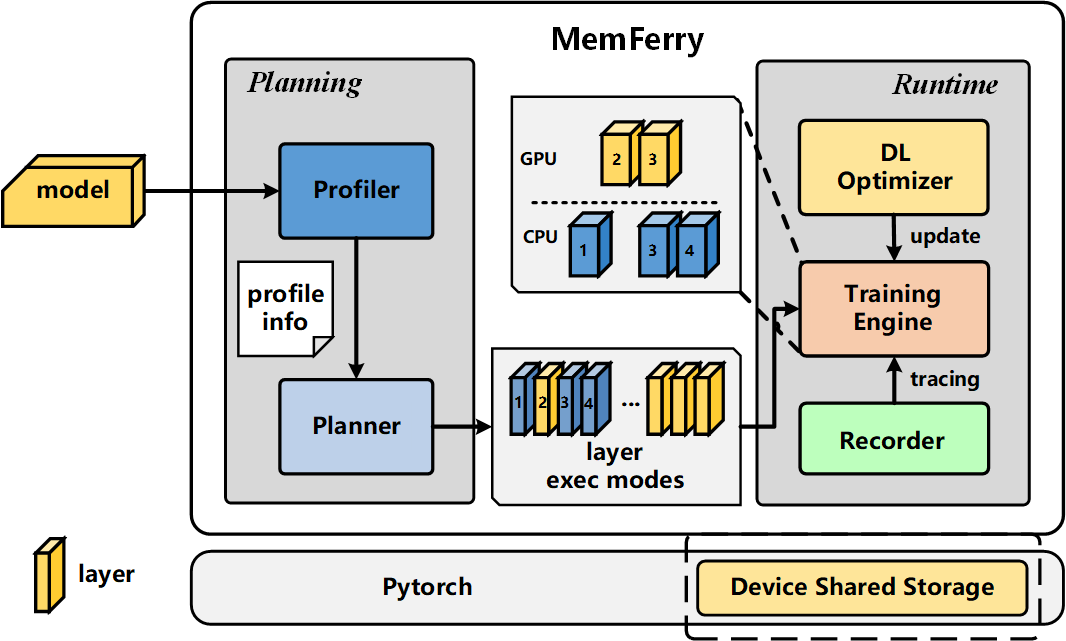}
	\centering
 \vspace{-0.3cm}
	\caption{Overview of \system System.}
	\label{fig:system_overview}
 \vspace{-0.5cm}
\end{figure}

 Fig.\ref{fig:system_overview} shows the overall architecture of \system, including the static planning components (\textit{Profiler} and \textit{Planner}), runtime memory management and execution logic components (\textit{Training Engine}, \textit{DL Optimizer}, \textit{Recorder}). When a new model training configuration initiates, \system first passes the training configuration into planning modules:
 
 \begin{itemize}
 \setlength{\itemsep}{0pt}
\setlength{\parsep}{0pt}
\setlength{\parskip}{0pt}
     \item \textbf{Profiler.} 
     The \textit{Profiler} uses different training configurations to evaluate the intra/inter-layer(s) computation times and the data transmission times.   
     
     \item \textbf{Planner.} The \textit{Planner} calls the scheduling algorithm based on the profiled information to determine the specific execution mode for each layer.
 \end{itemize}

Once the planning procedure finishes, \system Runtime modules start the model training. 

\begin{itemize}
\setlength{\itemsep}{0pt}
\setlength{\parsep}{0pt}
\setlength{\parskip}{0pt}
    \item \textbf{Training Engine.} The \textit{Training Engine} is responsible for memory management and data transmission. At the beginning of training, it distributes parameters to the corresponding devices and constructs logical and underlying storage models (including \textit{shadow model}). During the training process, it enforces the execution to behave as planned.

    \item \textbf{DL Optimizer.} 
    The \textit{DL Optimizer} performs parallel partial gradient transmission support (DMA parameters) and parameter update by user-specific training optimizer.

    \item \textbf{Recorder.} The \textit{Recorder} is a runtime monitor that helps to check the actually consumed time of each training process. 
\end{itemize}

In the rest of this section, we introduce the technical details of \system and its major features.

 \noindent \textbf{Workflow} 
 When a new model training configuration initiates, \system first passes the training configuration into \textit{Profiler}. The \textit{Profiler} uses different training configurations to evaluate the intra/inter-layer(s) computation times and the data transmission times. Then the \textit{Planner} calls the scheduling algorithm (See in Section \ref{sec:system design/training schedule}) based on the profiled information to determine the specific execution mode for each layer.

 Once the planning procedure finishes, \system Runtime modules start the model training. The \textit{Training Engine} is responsible for memory management and data transmission. At the beginning of training, it distributes parameters to the corresponding devices and constructs logical and underlying storage models (including \textit{shadow model}). During the training process, it schedules the execution to behave as planned. The \textit{DL Optimizer} performs parallel partial gradient transmission support (DHA parameters) and parameter update by user-specific training optimizer.

 \subsection{Hybrid Training Schedule}
 \label{sec:system design/training schedule}

\tcr{We hereby propose the layer-level scheduling algorithm of \system's scheduler for general foundation models. 
The design goal is to overlap parameter transmission with DHA computation as much as possible in FP and mitigate the adverse impact of enabling DHA in BP. We create three execution modes for the parameter of each layer: \textbf{GFGB} (GPU Forward GPU Backward), \textbf{DHA}, and 
\textbf{DFGB} (GPU DHA Forward GPU Backward). Let us take a layer of parameters as a unit. 
In GFGB, the parameters are transmitted from CPU memory to GPU memory before computation; in DHA, the parameters in CPU memory are computed by GPU SMs directly and no further transmission happens; in DFGB, the layer is transmitted to GPU memory after its DHA FP execution. DFGB helps to reduce BP time because the on-GPU BP is faster than the DHA BP (see in Section \ref{sec: motivation/GPU Computation with Host Memory}) so it can be regarded as a remedy of DHA for BP.} 

 \begin{figure}
	\includegraphics[scale=0.5]{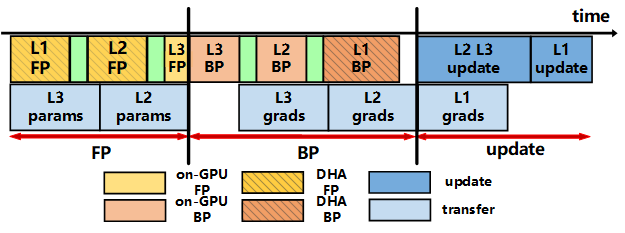}
	\centering
  \vspace{-0.3cm}
	\caption{Hybrid training schedule of \system.}

	\label{fig:hybrid training}
 \vspace{-0.6cm}
\end{figure}

Fig.\ref{fig:hybrid training} illustrates a training schedule example for a three-layer model. \tcr{We stipulate the execution modes of 1-3 layers to be DHA, DFGB and GFGB respectively. Due to DHA computation without any transmission, layer 1 and 2 start FP computation at the iteration beginning. At the same time, the parameter transmission of layer 3 also starts. When transmission of layer 3 is completed, \system load parameter of layer 2 from CPU to GPU for on-GPU BP, which avoids prolonging the BP time of layer 2. As we hide the gradient transmission with both BP and update, update does not wait for all the gradients to be transmitted.}

\tcr{To realize the above schedule, our first step is to profile the training information for \system's decision-making. 
During the profiling process, we evaluate the parameter transfer time, FP and BP computation time for each fine-grained layer. The profiling is executed only once, thus bringing negligible overhead. We summarize the profiled variable and their notations below.} Here, 
$L$: the number of layers; $l$: the layer id of model; $A_{l, li+1}$: computation between layer $l$ and $l{+}1$; $T_{trans}(l)$: parameter transfer time of layer $l$; $T_{FP}(l)$: on GPU forward time of layer $l$; $T_{BP}(l)$: on GPU backward time of layer $l$; $T_{FP}^{H}(l)$: DHA forward time of layer $l$; $T_{BP}^{H}(l)$: DHA backward time of layer $l$; $T_{FP}(A_{l,l+1})$: forward computation time $A_{l,l+1}$; $T_{BP}(A_{l,l+1})$: backward computation time of $A_{l,l+1}$; $S_{l}$: size of layer $l$.

\begin{comment}    
 \begin{table}
\caption{Profile Information}
  \label{table: profile infomation}
  \centering
  \begin{tabular}{|m{2cm}<{\centering}|m{5.5cm}<{\centering}|}
    \hline \textbf{symbol} & \textbf{description} \\
    \hline $L$ & the number of layers \\
    \hline $l$ & the layer id of model \\
    \hline $A_{l, li+1}$ & computation between layer $l$ and $l{+}1$\\
    \hline $T_{trans}(l)$ & parameter transfer time of layer $l$\\
    \hline $T_{FP}(l)$ & on GPU forward time of layer $l$\\
    \hline $T_{BP}(l)$ & on GPU backward time of layer $l$\\
    \hline $T_{FP}^{H}(l)$ & DHA forward time of layer $l$\\
    \hline $T_{BP}^{H}(l)$ & DHA backward time of layer $l$\\
    \hline $T_{FP}(A_{l,l+1})$ & forward computation time $A_{l,l+1}$\\
    \hline $T_{BP}(A_{l,l+1})$ & backward computation time of $A_{l,l+1}$\\
    % \hline $T_{GBP}(l)$ & gradient offload backward time  of layer $l$\\
    % \hline $T_{CBP}(l)$ & gradient on-GPU backward time of layer $l$\\
    
    % \hline $P_{l}$ & execution plan of layer $l$\\
    % \hline $P^G(l)$ & gradient offload plan of layer$l$\\
    \hline $S_{l}$ & size of layer $l$\\
    \hline
    % \hline $S_{req}$ & the requirement of space that set by user\\
  \end{tabular}
\end{table}
\end{comment}

\tcr{With the profiled information, we introduce our scheduling algorithm. We logically separate the entire training algorithm into the forward and backward algorithms.}

\noindent\textbf{Forward Algorithm (Alg. \ref{algorithm: FP process}).} Our goal is to minimize the overall execution time and hide transmission time as much as possible in computation. The execution plan is denoted as $P$ and the overall execution time as $T_{all}$. \tcr{We notice that computation can provide some overlapping time for communication. This overlapping time is called the reserved overlap time and denoted as $T_{res}$.} From the previous observations, we learn that loading time is often larger than computation time. Based on this, we make a simple criterion for judging each layer's execution mode. When the DHA computation time is less than on-GPU computation with loading time, using DHA can reserve more overlapping time for other layers to load parameters (i.e. $T_{res}$ increases) and reduce execution time. On the contrary, using GFGB mode is intuitively better.
%\vspace{-2cm}
% \renewcommand{\floatpagefraction}{0.60}

% \SetAlgoSkip{SkipBeforeAndAfter}
\begin{algorithm}[t]
\label{algorithm: FP process}
\SetKwFunction{FBack}{backtrack}
\SetKwFunction{FUpdate}{update}
    \caption{\system FP Algorithm}
    \KwInput{$T_{trans},\ T_{FP}, \ T_{FP}^{H},\  T_{FP}(A)$}
    \KwOutput{$P$}
        $T_{all},\ T_{res},\  tag \xleftarrow{} 0,\ 0,\ -1$\;
        \For{$l\  \xleftarrow{}\  1\  \KwTo\  L$}{
            \eIf{$ T_{FP}^{H}(l)\leq T_{FP}(l) + \max{(T_{trans}(l), 0})$}{
                $P_l \xleftarrow{} DHA$ \;
                % \textcolor{blue} {\Comment{DHA faster}}
                $update(T_{FP}^{H}(l),T_{FP}^{H}(l))$\;
            }{
                \eIf{$T_{trans}(l)<T_{res}$}{
                    \lIf{$tag = -1$} {$tag \xleftarrow{} l$}
                    $update(T_{FP}(l),T_{FP}(l) - T_{trans})$\;
                }{
                    $backtrack(l,tag,T_{all},T_{res})$\;
                }
                $P_l \xleftarrow{} GFGB$\; 
                % \textcolor{blue} {\Comment{naive on-GPU}}
            }
            $update(T_{FP}(A_{l,l+1}),T_{FP}(A_{l,l+1}))$\;
        }
        \Return $P,\  T_{res},\  tag$\;
        \BlankLine
        \Fn{\FBack{$l$,$bt$,$T_{all}^b$,$T_{res}^b$}}{
            $base \xleftarrow{} T_{all}^b+T_{trans}(l)+T_{FP}(l)-T_{res}^b$\;
            \For{$i\  \xleftarrow{}  bt\  \KwTo\  l$}{
                $gain\xleftarrow{}T_{trans}(i)+T_{FP}^{H}(i)-T_{FP}(i)$\;
                $T_{all}^{i} \xleftarrow{} \max(T_{trans}(l)-T_{res}^b-gain, 0)+T_{all}^b +(T_{FP}^{H}(i)-T_{FP}(i)) + T_{FP}(l)$ \;
                % \textcolor{blue}{\Comment{global time}}

                \If{$T_{all}^{i}<baseline$}{
                    $P_i,\  tag \xleftarrow{} DHA,\ i+1$ \;
                    % \textcolor{blue}{\Comment{confirm backtrack}}
                    $base,\  T_{res}^b \xleftarrow{} T_{all}^{i},\ T_{res}^b+gain$\;
                    \lIf{$T_{res}^b\geq T_{trans}(l)$}{
                        {\bf break}
                    }
                }
            }
            $update(base, \max(T_{res}^b-T_{trans}(l),0)+T_{FP}(l))$\;
        }
        \BlankLine
        \Fn{\FUpdate{$t_o$,$t_r$}}{
            $T_{all}, T_{res} \xleftarrow{} T_{all} + t_o,\  T_{res} + t_r $\;
        }
\end{algorithm}

\begin{figure}
\vspace{-0.5cm}
	\includegraphics[scale=0.43]{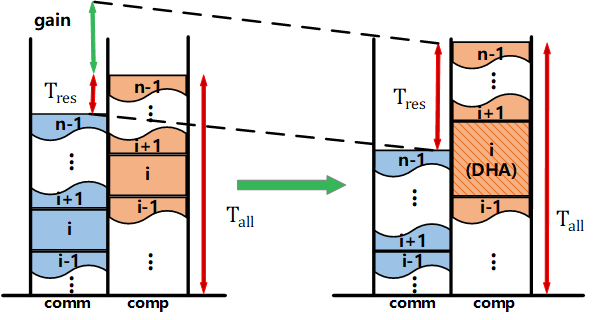}
	\centering
 \vspace{-0.3cm}
	\caption{Backtrack to turn execution mode of layer $i$ to DHA.}
	\label{fig:backtracking illustration}
 \vspace{-0.8cm}
\end{figure}

\tcr{As the above intuitive methods do not guarantee the optimality, we design a backtracking function.
A layer that is decided to execute using GFGB mode may have two cases. When the reserve time $T_{res}$ is greater than its loading time $T_{trans}$, it has sufficient time to load its parameters and then compute. When $T_{res}$ is insufficient to cover its loading time, we backtrack the previous GFGB layers and modify their execution mode to DHA mode, which provides more overlap time for the layer to load parameters.}

Figure \ref{fig:backtracking illustration} shows the effect on $T_{res}$ and $T_{all}$ when we modify a GFGB mode layer $i$ to DHA mode. After execution mode changing, $T_{res}$, $T_{all}$ will increase and we labeled them $T_{res}^i$, $T_{all}^i$. As in the figure, the gain of $T_{res}$ could be computed as $T_{res}^i-T_{res}=T_{all}+T_{FP}^H(i)-T_{FP}^H(i)+T_{trans}(i)$. In the backtracking process for a new GFGB layer $l$, after changing the execution mode of a previous layer $i$, if the $T_{trans}(i) - T_{res}^i-T_{res}$ is larger than $T_{all}^i-T_{all}$, then the mode change of layer $i$ successfully hides more parameter load time of layer $l$ without prolonging overall execution time.

\noindent\textbf{Backward Algorithm.} The PCIe communication is idle for a certain period between the end of parameter loading from CPU memory and the beginning of gradient transmission back to CPU memory. To fully utilize this idleness to reduce the negative impact of DHA computing, we present the DFGB execution mode that loads several layers of parameters (after FP) still in CPU memory into GPU memory. For these layers, their BP can be performed on GPU so as to reduce the time of obtaining the gradients. \tcr{We design a greedy algorithm to decide whether the layers are executed in DFGB mode.}

\begin{algorithm}
\label{algorithm: BP process}
    \caption{\system BP process}
    \KwInput{$T_{trans},\ T_{BP},\ T_{BP}^{H},\  T_{BP}(A),\ P,\  T_{res}, tag$}
    \KwOutput{$P$}
        $T_{res}\xleftarrow{}T_{res}+\sum^{L}_{l=tag+1}(T_{BP}(l)+T_{BP}(A_{l,l+1}))$\;
        $load\_dict \xleftarrow{} \varnothing$\;
        \For{$l\  \xleftarrow{}\  L\  \KwTo\  1$}{
            \If{$P_l=1\ and\ T_{trans}(l)+T_{BP}(l)<T_{BP}^{H}(l)$}{
                $load\_dict.add(T_{BP}^{H}(l)-T_{trans}(l)-T_{BP}(l),\ l)$\;
            }
        }
        
        $SortbyKeyDesc(T_{res})$\;
        
        \ForAll{$g,\ l\  \in\ load\_dict$}{
            \If{$T_{res}>T_{trans}(l)$}{
                $P_l,\  T_{res} \xleftarrow{} DFGB,\ T_{res}-T_{trans}(l)$ \;
                % \textcolor{blue}{\Comment{on-GPU BP}}
            }
        }
        \Return $P$\;
\end{algorithm}

\subsection{Memory Management}
\label{sec:system design/memory management}

\tcr{The hybrid training requires \system to be capable of computing data stored in both the GPU memory and the CPU memory.} This signifies that \system will stipulate the storage space for different partitions of the model and their respective access patterns. We then design an abstract model named \textit{shadow model} with a set of merits: \textit{shared memory of GPU and CPU}, \textit{minimal memory copy} and \textit{compatible to ML frameworks}. We next explain where the model partitions are stored and how they interact with the shadow model. 

\begin{figure}%[htbp]
	\includegraphics[scale=0.3]{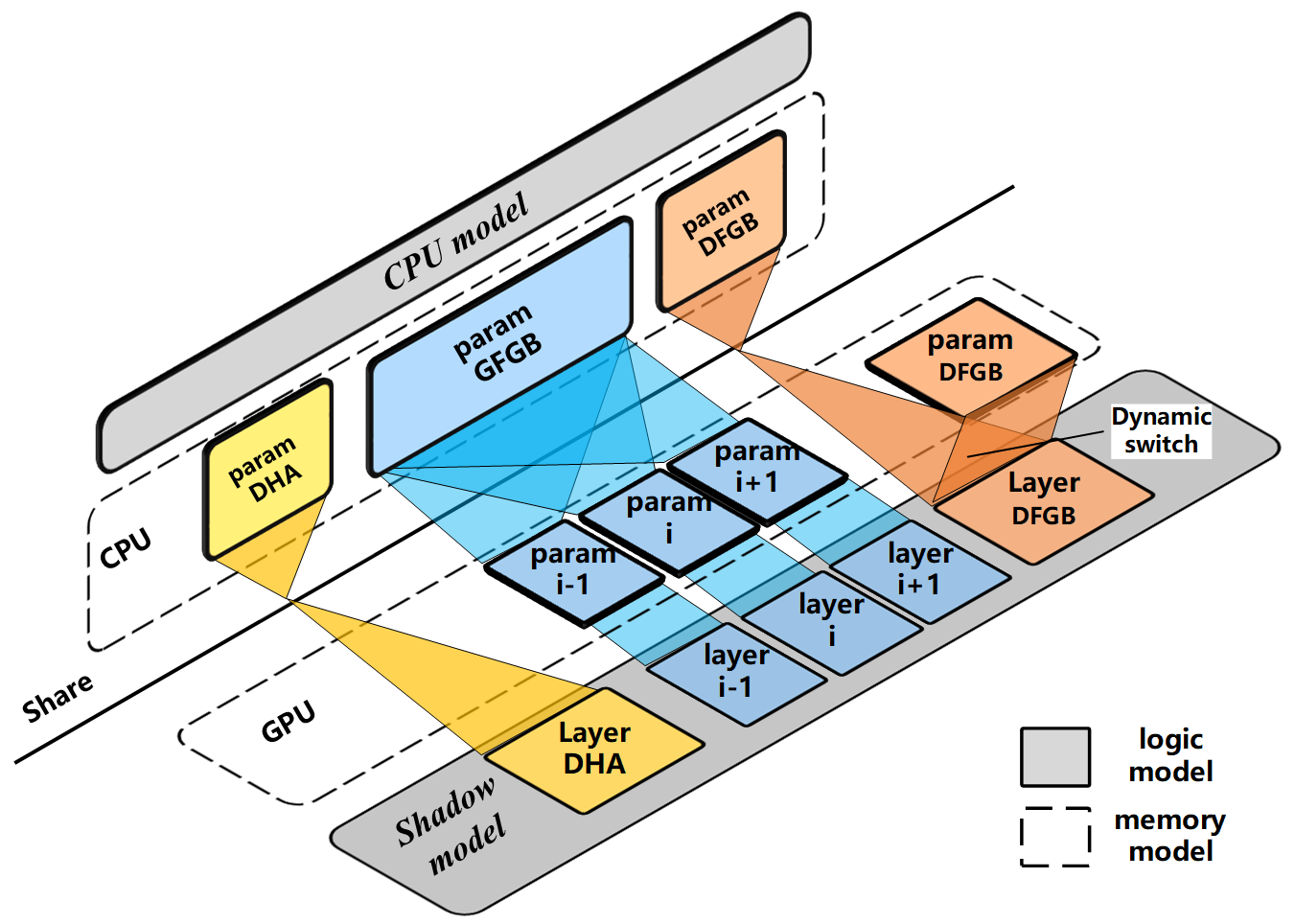}
	\centering
 \vspace{-0.3cm}
	\caption{The shadow memory management in \system.}
	\label{fig:memory management}
 \vspace{-0.6cm}
\end{figure}

We firstly separate the models of the logical level and the underlying storage level. In \textit{Training Engine}, a logical CPU model (for updates), a logical GPU model (for FP\&BP) and CPU-GPU parameters storage are maintained, as the Figure \ref{fig:memory management} shows.
In the CPU memory, the model parameters are partitioned into three parts corresponding to the three execution modes in Section \ref{sec:motivation/exection schedule}: 1) GFGB parameters that will be transmitted to the GPU memory; 2) DHA parameters that will be computed in situ by GPU; 3) DFGB parameters that will be computed by GPU first and be transmitted to GPU memory later on. To maximize the parallel computing power of the CPU SIMD, we aggregate the GFGB parameters into a long parameter vector. Besides, the CPU memory also hosts the gradient and the optimizer state just like ZeRO-Offload. The GPU runtime memory stores the GFGB parameters buffer and the DFGB parameters buffer in line with the corresponding parameters on CPU memory.

The logical GPU model used in FP\&BP is called the \textbf{\textit{shadow model}}. \system adopts a dynamic mapping mechanism for the underlying storage of shadow model based on the execution plan. The parameters of the shadow model are not necessarily located in the GPU memory but may also be mapped from the CPU parameters. For example, the parameters executed by the GFGB mode are projected and transmitted from the long parameter vector to GPU memory; for the parameters executed by DHA mode, \system directly delivers their storage pointers to the shadow model, which could access them directly. \tcr{As shown in Figure \ref{fig:memory management}, We use three colors to differentiate the execution modes of GFGB (\textit{blue}), DHA (\textit{yellow}) and DFGB (\textit{orange}). The DHA parameters are directly mapped from CPU memory to the shadow model in GPU.} The GFGB parameters are coordinated to a large contiguous area in CPU memory, and mapped separately to GPU memory layer by layer when being loaded. The DFGB parameters have both the DHA storage and the GPU memory storage, whose pointer can be switched before BP starts.

The shadow model abstraction brings two advantages: 1) \textbf{Reducing space allocation and memory copy.} \tcr{by unified storage of logical parameters. We directly share some of the DHA parameters of the CPU model with the GPU model, avoiding redundancy and copying in the memory buffer.} 2) \textbf{Facilitating dynamic storage switching in DFGB.} \system dynamically modifies the mapping of the physical storage of the DFGB layer after FP and change it back after BP without additional burden or affecting the learning procedure.

\subsection{Gradient Offloading}
\label{sec:system design/gradient offload}

A crucial question regarding DHA-enabled training is whether the GPU memory requirement can be further relieved. With the ever-growing model size, \system may still run out of GPU memory. To achieve the flexible usage of GPU memory and avoid violating the memory limit, we present \systemMem based on the aforementioned scheduling algorithms and observations.

\tcr{The GPU content of \system includes the partial model parameters, the gradients generated by BP and intermediate activation. Further offloading model parameters will significantly decrease the training speed of layers that are not suitable for DHA. Offloading intermediate activation values involves a large number of reads and writes between GPU and CPU through slow PCIe. Therefore, \systemMem chooses to offload the gradients. Specifically, \systemMem designates a CPU memory as DHA memory, mapping it to the gradient storage of the shadow model. During the BP process, \systemMem will directly generate the gradient in the CPU memory, avoiding the transferring and copying of GPU memory.} 

However, we do not expect gradient offloading to bring a significant side effect to training speed, and the amount of memory offloaded should be determined based on the needs of users. \systemMem does not completely offload the gradients of the entire model but the gradients of partial parameters while meeting the user's memory needs. In Section \ref{sec: motivation/GPU Computation with Host Memory}, we find that different model parameters have different performances on gradient offloading. \tcr{Based on this phenomenon, \systemMem automatically determines the layers for gradient offload to ensure maximum training speed under the user-specified memory requirement.}

\textbf{Gradient offload algorithm.} We design a gradient offload algorithm based on dynamic programming. In this algorithm, users need to specify the total percentage $S_{req}$ of gradient on GPU (named \textit{reserve factor}), and the algorithm outputs the specific layers to offload as the gradient offload plan $P^{GO}$. We denote the normal on-GPU BP time for each layer as $T_{CBP}$, while the BP time after offloading the gradient by DHA is $T_{GBP}$. We default to offload gradients for all layers at the beginning and then judge the gain $T_{GBP}-T_{CBP}$ in time when adjusting a layer's gradient to on-GPU mode. If we define the problem $A_{l}$ which means the maximum gain offload strategy of layer $1$ to $l$. Then we could find that for a specific layer $i$, $A_{i}$ is up to subproblem $A_{i-1}$. if layer $i$ is choose to keep gradient offloading then $A_{i}=A_{i-1}$, else if layer $i$ turn gradient to GPU then $A_{i}=A_{i-1}+T_{GBP}(i)-T_{CBP}(i)$. \tcr{Based on this optimal substructure, we construct a dynamic programming table $D_{l,s}$ from two dimensions: layer id and space. The second axis of the table represents the size of the space occupied by the current plan, which is used to limit overall space. The content of the table is the execution time gain of the current plan (i.e. $A$). The recursion formula of $D$ is given by:}
$$
D_{i,j} = \max(D_{i-1,j},\  D_{i-1,j-S_{i}}+T_{GBP}(i)-T_{CBP}(i)).
$$

By using dynamic programming to identify the gradient offloading plan that has the least execution time, \systemMem ensures the use of gradient offload meets the memory requirement with the least side effect. This allows users to adjust their memory requirements based on the actual memory size even after offloading some parameters and optimizer states during the various training processes. \tcr{When the specified gradient is fully offloaded, the GPU memory required for \systemMem is only the size of partial parameters and activations.}

\subsection{Adapting to modern scale-up domains}
\label{sec:design/scale-up}
The emerging scale-up architectures provide new opportunities to further accelerate the data movement in \system. The original \system is designed for conventional CPU--GPU interconnections, where parameters and gradients are directly transferred between CPU memory and the target GPU through PCIe. In modern scale-up domains, however, accelerators are interconnected with much higher bandwidth and richer connectivity than CPU--accelerator links. For example, in our target CloudMatrix configuration, each CPU provides five UB links while an NPU exposes seven high-speed UB ports. Therefore, directly transferring data from CPU$_0$ to NPU$_0$ may be bounded by the limited CPU--NPU link bandwidth, while substantial bandwidth among NPUs remains unused. This mismatch is particularly undesirable for \system because large amounts of parameters and gradients are repeatedly transferred between CPU and accelerator memories. To exploit the additional bandwidth provided by scale-up fabrics, we extend \system with scale-up-assisted data transfer and an adaptive data dispatcher.

 \begin{figure}%[htbp]
	\includegraphics[scale=0.5]{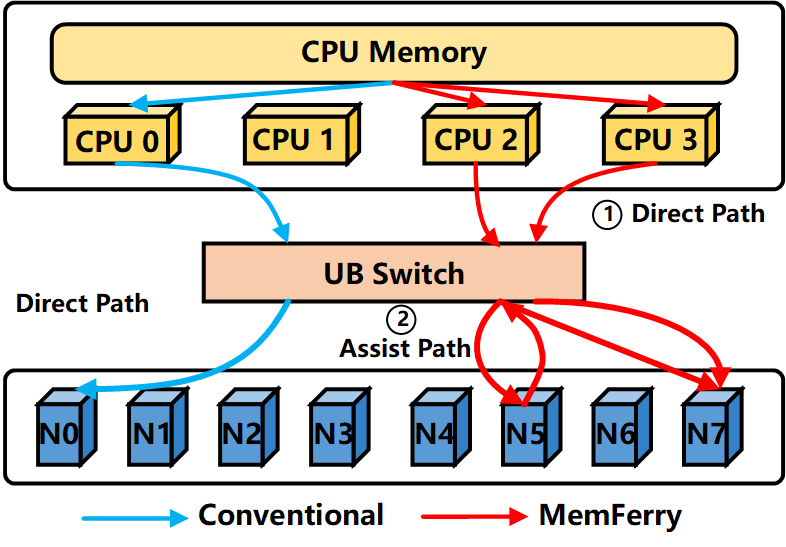}
	\centering
 \vspace{-0.3cm}
	\caption{Scale-up assisted data transfer.}
	\label{fig:assist transfer}
 \vspace{-0.6cm}
\end{figure}

\noindent \textbf{Scale-up assistent data transfer.} 
Consider transferring a tensor from CPU$_0$ memory to NPU$_0$. Besides the conventional direct path CPU$_0 \rightarrow$ NPU$_0$, \system introduces an assisted path CPU$_0 \rightarrow$ NPU$_1 \rightarrow$ NPU$_0$, where NPU$_1$ temporarily serves as a forwarding device. As shown in Figure \ref{fig:assist transfer}, the two paths transmit different partitions of the tensor concurrently. Since the NPU--NPU bandwidth in a scale-up fabric is considerably higher than the CPU--NPU bandwidth, forwarding the data from NPU$_1$ to NPU$_0$ introduces relatively small transmission overhead for large tensors. More importantly, the assisted path utilizes another CPU egress link and otherwise idle NPU interconnect bandwidth, thereby increasing the effective bandwidth from CPU memory to the target NPU. This design can naturally use multiple assisting NPUs when available. Given a set of assisting NPUs $\mathcal{A}$, the additional transfer capacity is approximately bounded by 
$$ 
\sum_{i \in \mathcal{A}} \min\left(B_{C \rightarrow i}, B_{i \rightarrow 0}\right), 
$$
subject to the CPU memory bandwidth and UB fabric contention. In this way, \system converts the abundant NPU--NPU connectivity in scale-up domains into additional bandwidth for offloaded training data movement.

\noindent \textbf{Adaptive data dispatcher.}
The assisted path provides higher aggregate bandwidth but introduces an additional forwarding hop, making it less attractive for small transfers whose performance is dominated by latency. MemFerry therefore dynamically partitions each transfer according to the latency and bandwidth of the two paths. Consider a tensor of size $S$, where $x$ bytes are assigned to the direct path and $S-x$ bytes to the assisted path. Denoting their startup latencies and effective bandwidths by $(L_d, B_d)$ and $(L_a, B_a)$, respectively, the transfer time is 
$$ T(x) = \max\left( L_d + \frac{x}{B_d}, L_a + \frac{S-x}{B_a} \right). $$
MemFerry chooses
$$ x^{*} = \arg\min_{0 \leq x \leq S} T(x). $$
When both paths are used, their completion times are balanced, yielding 
$$ x^{*} = \left[ \frac{ B_d \left(S + B_a(L_a-L_d)\right) }{ B_d+B_a } \right]_{0}^{S}, $$
where $[\cdot]_{0}^{S}$ clips the value into $[0,S]$. Consequently, when 
$$ S \leq B_d(L_a-L_d), $$
the entire tensor is transferred through the low-latency direct path. As $S$ increases, more data is assigned to the assisted path, and its fraction asymptotically approaches
$$ \frac{B_a}{B_d+B_a}. $$
MemFerry profiles $(L_d,B_d,L_a,B_a)$ for the underlying topology and applies the resulting allocation to parameter and gradient transfers, preserving low latency for small tensors while exploiting the higher aggregate bandwidth of scale-up fabrics for large transfers.

\section{Implementation}
\label{sec:implementation}

We elaborate on the implementation of \system as below. 

{\color{black}
\textbf{MemFerry Memory Management.} (1) \emph{Memory sharing.} \system needs a programming interface to share the DMA memory between the CPU model and the GPU shadow model. The CPU memory pointers can be shared through this interface and used by the GPU SMs. We implement a memory allocator class, \textit{THCCachingCUDAHostSharedAllocator}, in PyTorch that returns the CPU memory address of the corresponding DMA tensor allocated in the GPU. Through this memory allocator, both the GPU shadow model and the CPU model can access the same physical address in CPU memory and share the tensor located in this area.

(2) \emph{DFGB memory switching.} The DFGB mechanism requires the switching of parameters between DMA tensors and GPU HBM tensors at runtime. Because the automatic differentiation of PyTorch records related computation operations or gradient functions of the parameter tensors, directly replacing parameter tensors in the logic layer may lead to malfunction in BP. To address this problem, \system replaces the \textit{storage} field of the tensor. Both the DHA tensor and CUDA tensor consist of \textit{storage} (\textit{storage} class includes memory pointers and access formats) and other description fields. So the parameter switching could be achieved by switching the \textit{storage} reference (an 8-bit field) of parameter tensors. In this way, we complete the implementation of DFGB parameter switching without any data copying or computation graph modification.

}

\begin{figure*}
\centering
  \subfigure[Bert-base]{
  \label{fig:evaluation/time of bert base}
  \begin{minipage}{.23\textwidth}
  \centering
    \includegraphics[scale=0.24]{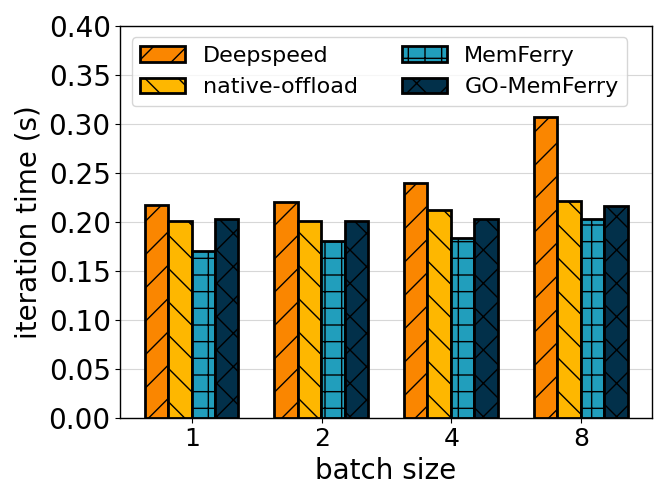}
  \end{minipage}
  }
  \subfigure[Roberta]{
  \label{fig:evaluation/time of roberta}
  \begin{minipage}{.23\textwidth}
  \centering
    \includegraphics[scale=0.24]{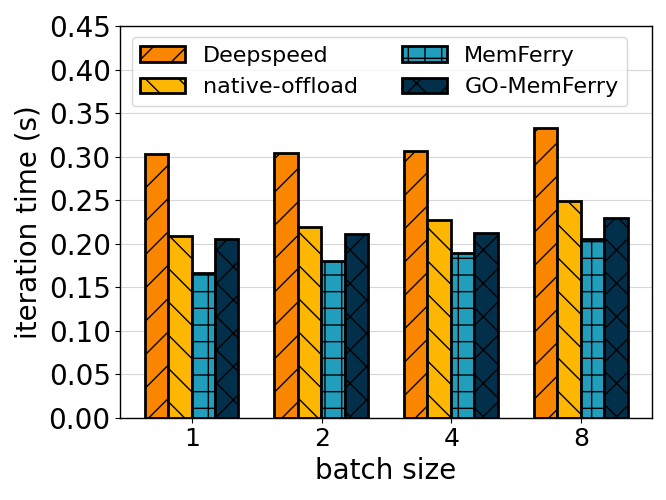}
  \end{minipage}
  }
  \subfigure[Bert-large]{
  \label{fig:evaluation/time of bert large}
  \begin{minipage}{.23\textwidth}
  \centering
    \includegraphics[scale=0.24]{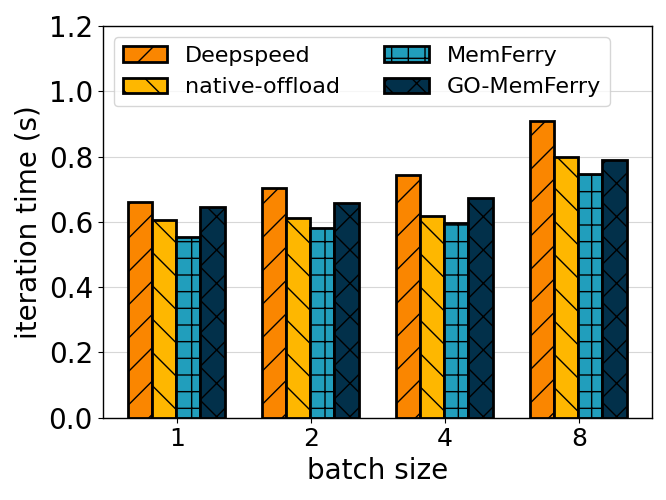}
  \end{minipage}
  }
  \subfigure[GPT-2]{
  \label{fig:evaluation/time of GPT2}
  \begin{minipage}{.23\textwidth}
  \centering
    \includegraphics[scale=0.24]{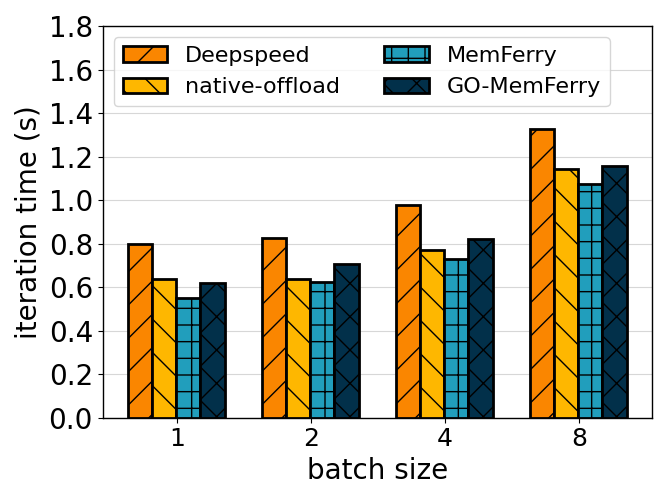}
  \end{minipage}
  }
  \vspace{-0.3cm}
   \caption{Overall iteration time.}
  \label{fig:evaluation/training time} 
\vspace{-0.5cm}
\end{figure*}

\begin{figure}
\centering
% \hspace{-0.4cm}
  \subfigure[Average GPU utilization]{
  \label{fig:evalution/GPU util}
  \begin{minipage}{.21\textwidth}
  \centering
  \hspace{-0.3cm}
    \includegraphics[scale=0.24]{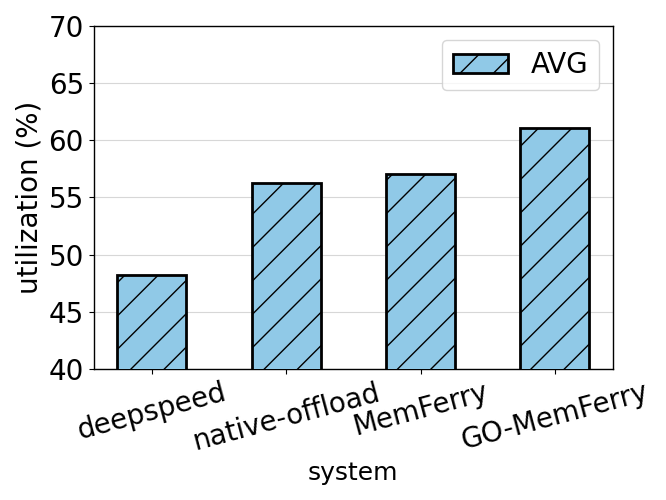}
  \end{minipage}
  }\hspace{0.4cm}
  \subfigure[Reserve factors]{
  \label{fig:evalution/time of GO_system}
  \begin{minipage}{.21\textwidth}
  \centering
    \includegraphics[scale=0.24]{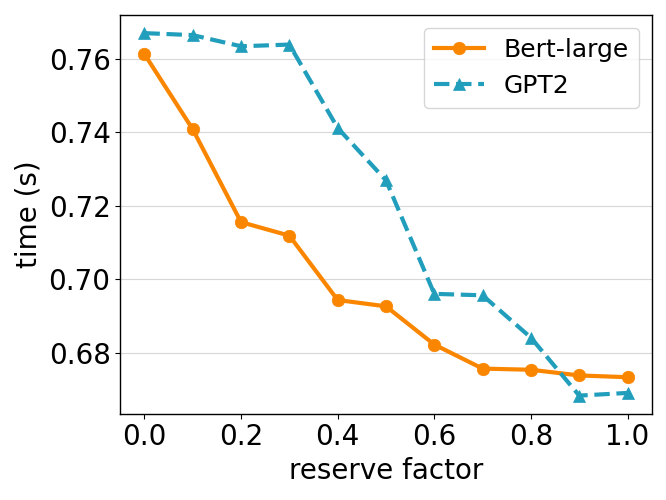}
  \end{minipage}
  }
  \vspace{-0.3cm}
   \caption{Utilization and reserve factor of \systemMem.}
  \label{fig:evaluation/mix figure} 
\vspace{-0.4cm}
\end{figure}

\begin{figure}%[htbp]
	\includegraphics[scale=0.25]{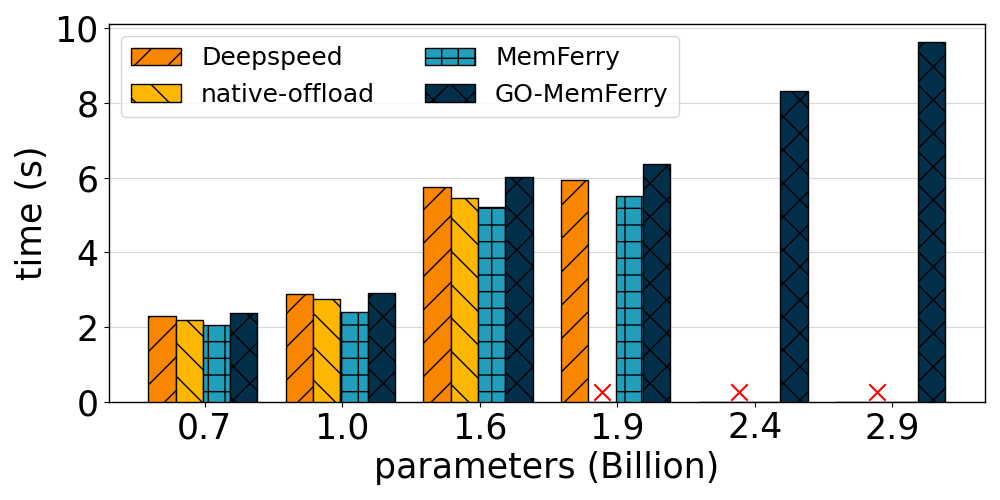}
	\centering
 \vspace{-0.5cm}
	\caption{Training performance on Nvidia V100 GPU.}
	\label{fig:evalution/V100 test}
 \vspace{-0.5cm}
\end{figure}

\begin{figure}
\centering
% \hspace{-0.4cm}
  \subfigure[Bert-base]{
  \label{fig:evalution/GO_system bert}
  \begin{minipage}{.21\textwidth}
  \centering
  \hspace{-0.3cm}
    \includegraphics[scale=0.24]{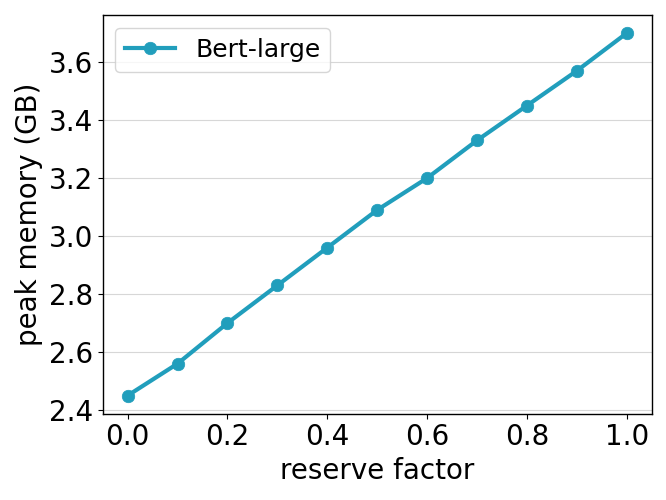}
  \end{minipage}
  }\hspace{0.4cm}
  \subfigure[GPT2]{
  \label{fig:evalution/GO_system gpt}
  \begin{minipage}{.21\textwidth}
  \centering
    \includegraphics[scale=0.24]{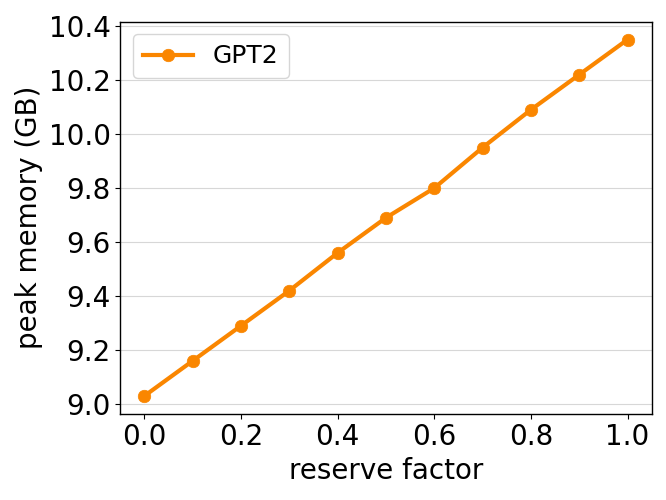}
  \end{minipage}
  }
  \vspace{-0.3cm}
   \caption{Memory usage of \systemMem during training}
  \label{fig:evalution/memory of GO_system} 
\vspace{-0.5cm}
\end{figure}

\begin{figure*}
\centering
  \subfigure[Bert-base (110M)]{
  \label{fig:evaluation/mem of bert base}
  \begin{minipage}{.23\textwidth}
  \centering
    \includegraphics[scale=0.24]{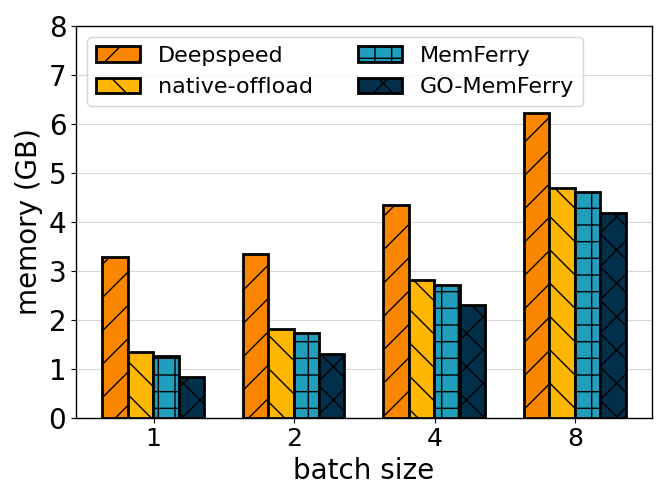}
  \end{minipage}
  }
  \subfigure[Roberta (110M)]{
  \label{fig:evaluation/mem of roberta}
  \begin{minipage}{.23\textwidth}
  \centering
    \includegraphics[scale=0.24]{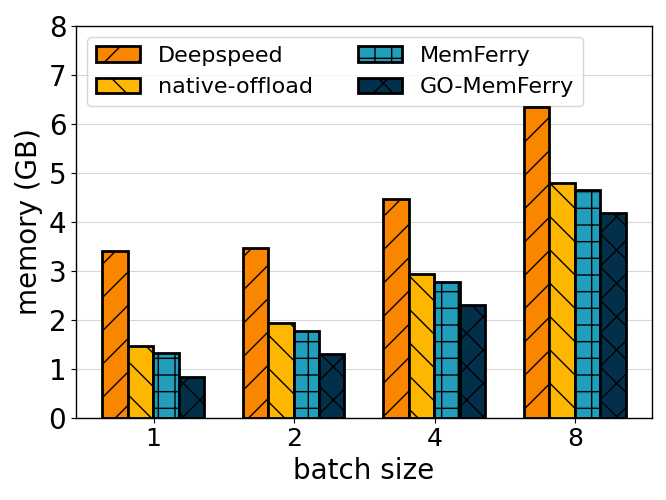}
  \end{minipage}
  }
  \subfigure[Bert-large (340M)]{
  \label{fig:evaluation/mem of bert large}
  \begin{minipage}{.23\textwidth}
  \centering
    \includegraphics[scale=0.24]{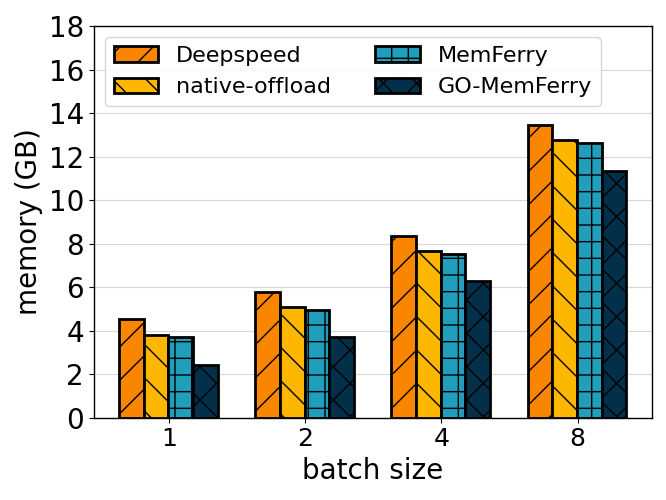}
  \end{minipage}
  }
  \subfigure[GPT-2 (355M)]{
  \label{fig:evaluation/mem of GPT2}
  \begin{minipage}{.23\textwidth}
  \centering
    \includegraphics[scale=0.24]{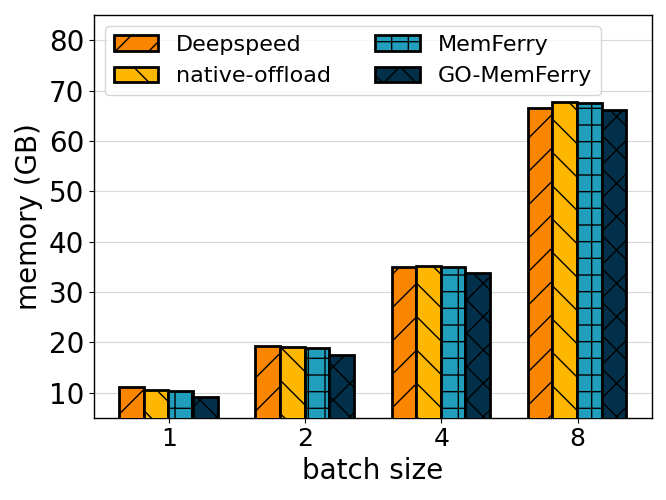}
  \end{minipage}
  }
  \vspace{-0.3cm}
   \caption{Maximum memory usage.}
  \label{fig:evaluation/training memory usage} 
  \vspace{-0.5cm}
\end{figure*}

\textbf{MemFerry Tensor Scheduling.} 
{\color{black} 
\system utilizes the asynchronous execution mechanism in PyTorch to schedule tensors. At the beginning of data transfer, we apply the \textit{non-blocking} option to the transmission function (such as \textit{copy\_}), so the CPU does not have to forcibly wait for the GPU to complete the transmission. After transmission on the GPU starts, CPU can start the corresponding computation, thereby achieving parallel execution of computation and transmission. When execution to the scheduling boundary in Sec.\ref{sec:system design/training schedule}, we added a mandatory synchronization barrier to ensure the consistency of transmitting data. 
}

\textbf{ScaleUp-MemFerry Implementation.}
We extend the Training Engine with a direct path and an assisted
path for CPU-to-NPU data movement. During profiling, MemFerry
measures the startup latency and effective bandwidth of both paths
for the underlying UB topology. The adaptive dispatcher computes
the data partition according to Sec.\ref{sec:design/scale-up} and launches the two
transfers concurrently. A synchronization is inserted before the
transferred tensor is consumed by the target NPU.

\section{Performance Evaluation}
\label{sec:evaluation}

\subsection{Experimental Setup}
\label{sec:evaluation/Setup}
% configuration, objects, workloads, metrics
{\color{black}
\textbf{Testbed and DNN Models.} Our evaluation is conducted on two cloud instances, \tcr{one equipped with $128$ CPU cores and 8 Nvidia A100 GPUs of 80$GB$ memory}, the other equipped with $12$ CPU cores and an Nvidia V100 GPU of 32$GB$ memory. Both of them are equipped with PCIe 3.0. The software we use includes Pytorch-1.9.0, Python-3.6 and CUDA-11.7. 
The foundation models in our experiments include  
Bert-base/Bert-large \cite{DBLP:conf/nips/bert}, Roberta \cite{DBLP:journals/corr/roberta}, GPT-2-medium \cite{gpt2} models of transformers \cite{DBLP:conf/emnlp/transformers} library and transformerXL \cite{DBLP:conf/acl/transformerXL}. In transformerXL, we manually configure a set of models of different sizes \tcr{within the range $0.7B$-$6.7B$ in order to evaluate} system performance in a controlled manner. 
The embedding layer size ranges from $32768$ to $130528$, the hidden layer size is from $2048$ to $4096$ and the sequence length is from $378$ to $4096$.
The scale-up evaluation is conducted separately on a CloudMatrix384 compute node and is described in Sec.\ref{sec: eva/ Scaling to scale-up domain}.

\textbf{Baselines and Metrics.} Our baseline systems include DeepSpeed implementation of ZeRO-Offload (DeepSpeed), native ZeRO-Offload (Native-Offload), \system and \systemMem. The implementation methods influence the performance of ZeRO-Offload and \system. In \system, we do not destroy the gradients on GPU after they have been used while DeepSpeed does so. This brings considerable latency for DeepSpeed to dynamically construct and deconstruct the gradient buffer. For fair comparison, we implement a native ZeRO-Offload that is faithful to its original design. 
% \systemMem can also be deemed as the more flexible \system in which the layers of gradients can be strategically removed from GPU memory. 
Note that for \systemMem, the default configuration is to completely offload the gradients via DHA. The performance metrics considered in our evaluation are threefold: \emph{per-iteration time}, \emph{GPU utilization}, and \emph{GPU memory usage}. 
For the scale-up evaluation in Sec.\ref{sec: eva/ Scaling to scale-up domain}, we additionally report CPU-to-NPU transfer bandwidth.

}

\subsection{Training Efficiency of \system}
\label{sec:evaluation efficiency}

{\color{black}

% \textbf{\system improves the training speed of various models.} 
\textbf{Speedup over various DNNs.}
We first evaluate the training speed of \system with different large language models in Figure \ref{fig:evaluation/training time}. Overall, the per-iteration time is consistently the highest in DeepSpeed, while it is always the shortest in \system. In all the models training we conducted, \system has a $10\%$ to $40\%$ improvement in training speed compared to DeepSpeed and Native-Offload. 
% \tcr{\systemMem has a longer training latency than \system because \systemMem uses DHA write for gradient computation during BP, which causes delays.}
When using the two smaller models (\textit{Bert-base}, \textit{Roberta}), the training speed of \system improves by $21.7\%$ to $68.7\%$ compared to the baselines.
As for larger models (\textit{Bert-large}, \textit{GPT-2}), the training speedup is reduced to $7.1\%$ till $32.1\%$. The main reason is that they usually require large inputs such that the computation time accounts for a larger proportion of the overall training period. 
}

% {\color{black}
\textbf{GPU utilization improvement.}
The GPU utilization when training \textit{GPT-2} is illustrated in Figure \ref{fig:evalution/GPU util}. 
 % using different systems
% The batch size is set to $8$ in all the experiments by default, and other batch sizes exhibit similar phenomena but are omitted here. 
\system's GPU utilization is $18.5\%$ higher than DeepSpeed and $1.5\%$ higher than Native-Offload. 
This utilization improvement originates from \system reducing the GPU idle time during communication and improving the utilization of GPU SMs through parallel DHA computation and transmission. 
Native-Offload has higher utilization than DeepSpeed because the latter needs time to allocate and reclaim gradient memory.
\tcr{\systemMem has the highest utilization ($27\%$ higher than DeepSpeed), as it parallel executes gradient computation and transmission.}

%\vspace{-0.1cm}
\subsection{Training Memory Usage}
\label{sec:evaluation/memory}

\textbf{Larger model size limitation.}
To demonstrate the training performance of \system and \systemMem under limited training resources, we conduct LLM training on the V100 GPU machine. In Figure \ref{fig:evalution/V100 test}, we configure models of different sizes with a training batch size of $1$
% and the data sequence length is $4096$. 
\system is still the fastest in all training configurations, $1.12\times$ faster than the baseline. When \systemMem fully offload gradients in this figure, the training speed is the slowest, but only $3.4\%$ lower than the baseline. This is because the lower PCIe bandwidth severely affects the write operations of DHA gradients. As the model size increases, Native-Offload, DeepSpeed, and \system all experience memory overflow. After the model is larger than $2.4B$, only \systemMem can train normally. Overall, \systemMem can support models that are $1.52\times$ larger than DeepSpeed ZeRO-Offload with limited memory.

{\color{black}

% \textbf{\systemMem significantly reduces GPU memory usage.}
\textbf{Memory usage reduction over various DNNs.}
\tcr{We then study the \emph{peak} GPU memory usage in an entire training iteration.} In Figure \ref{fig:evaluation/training memory usage},
\system reduces GPU memory by average $40.1\%$ compared with DeepSpeed when training smaller language models. \tcr{Compared to baselines, \system uses DHA computing to move some parameters from the GPU to the CPU.} \systemMem has a memory reduction of $17.6\%$ even compared to \system, as the gradients are completely removed from the GPU memory.  
When training  \textit{GPT-2}, \system saves the GPU memory by an average of $1.4\%$.
On average, \system and \systemMem reduce GPU memory by $ 25.4\%$ and $ 41.6\%$.

}

\begin{figure}
\centering
% \hspace{-0.4cm}
  \subfigure[Iteration time]{
  \label{fig:evalution/llm time}
  \begin{minipage}{.21\textwidth}
  \centering
  \hspace{-0.35cm}
    \includegraphics[scale=0.24]{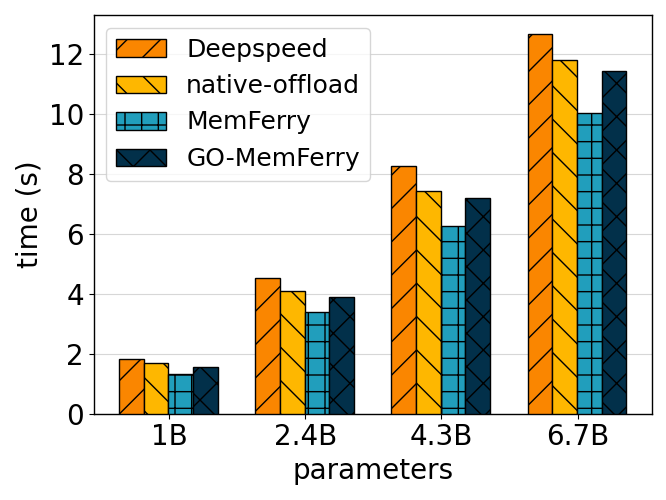}
  \end{minipage}
  }\hspace{0.4cm}
  \subfigure[Memory usage]{
  \label{fig:evalution/llm mem}
  \begin{minipage}{.21\textwidth}
  \centering
    \includegraphics[scale=0.24]{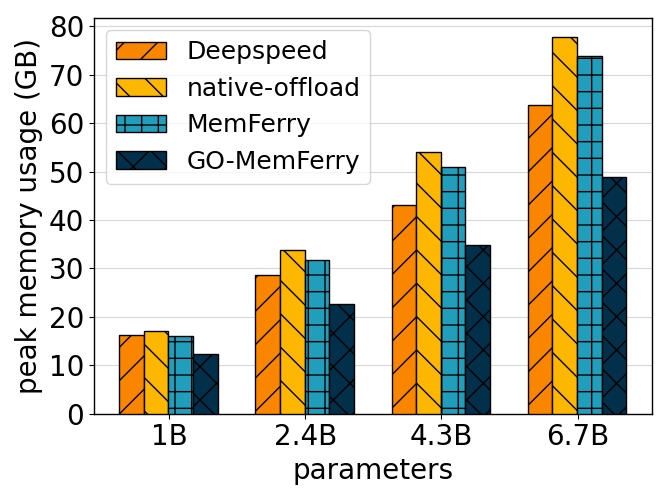}
  \end{minipage}
  }
  \vspace{-0.3cm}
   \caption{Training on large language models.}
  \label{fig:evaluation/large models} 
\vspace{-0.5cm}
\end{figure}

\begin{figure}
\centering
% \hspace{-0.4cm}
  \subfigure[Scalability to number of GPUs (batch size = 16)]{
  \label{fig:evalution/number of GPUs}
  \begin{minipage}{.21\textwidth}
  \centering
  \hspace{-0.3cm}
    \includegraphics[scale=0.24]{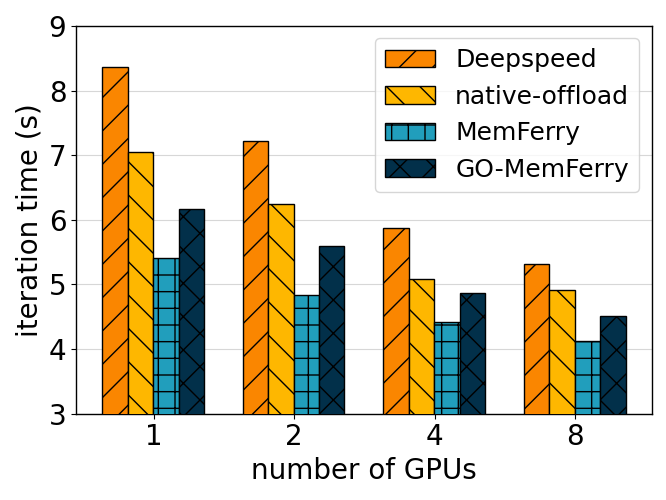}
  \end{minipage}
  }\hspace{0.4cm}
  \subfigure[Scalability to batch size (8 GPUs)]{
  \label{fig:evalution/batch size GPU}
  \begin{minipage}{.21\textwidth}
  \centering
    \includegraphics[scale=0.24]{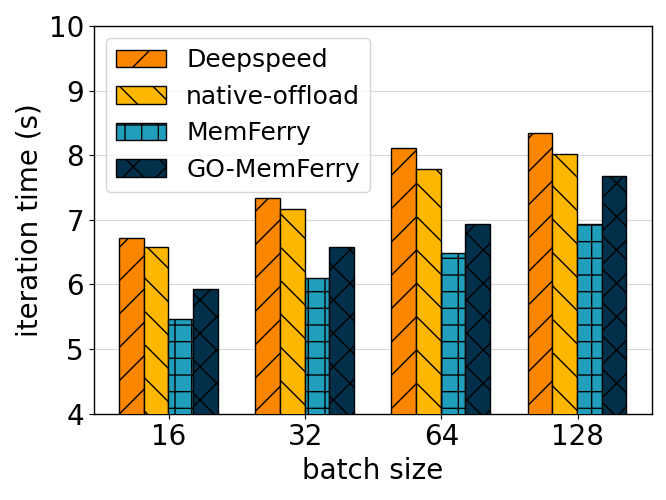}
  \end{minipage}
  }
  \vspace{-0.3cm}
   \caption{Performance in multi-GPU training.}
  \label{fig:evaluation/mix figure} 
\vspace{-0.6cm}
\end{figure}

{\color{black}

\textbf{Flexibility of \systemMem.} As shown in Fig.\ref{fig:evalution/memory of GO_system}, we first measure the GPU memory usage with different reserve factors. In the training of both Bert-base and GPT2, the memory usage of \systemMem is proportional to the reserve factor, which means that ML engineers can flexibly determine the gradient offloading scheme using  \systemMem, given their GPU memory budgets. When offloading the gradients completely, \systemMem saves an average of $20.9\%$ GPU memory compared to \system.
While offloading gradients within the user-specified memory size based on the reserve factor, \systemMem minimizes the degradation in the training speed. 
In Figure \ref{fig:evalution/time of GO_system}, we measure the iteration time of \systemMem when the gradient reserve factor (larger means keep more gradients on GPU) varies. \tcr{As the reserve factor increases, iteration time gradually decreases due to less DHA gradient write. The curves are convex, indicating that \systemMem's gradient offload algorithm offloads the gradients with the smallest performance loss to meet users' needs.}
When the reserve factor increases, \systemMem leans to \system, resulting in a higher training speed.

% more gradients being generated on GPU which avoids slower

\subsection{Scalability of \system}

\textbf{Scalability to model size.}
Figure \ref{fig:evalution/llm time} shows the training speeds of \textit{TransformerXL} with different parameter sizes where the batch size is set to $8$. \system always possesses the highest training speed, $1.32\times$ faster than DeepSpeed on average. As the model size increases, the speed advantage of \system decreases. When training the $1B$ model, \system has a training speed improvement of $1.37\times$ compared to DeepSpeed, while training the $6.7B$ model the improvement is $1.26\times$. This is because when the model is larger, the computation time during the training process also significantly increases. Figure \ref{fig:evalution/llm mem} shows the memory usage. Overall, \systemMem uses the least space, with an average reduction of $21.6\%$ compared to DeepSpeed. As the model size increases, the maximum memory savings of \systemMem remain between $18\%$ and $22\%$. indicating that \systemMem can reduce the memory requirements for large-scale model training.

}

\textbf{Scalability to multiple GPUs.} MemFerry supports data parallel training that ``aggregates'' the computing power of multiple GPUs (using Pytorch  DataParallel). 
%We then study multi-GPU training performance by integrating \system with Pytorch DataParallel. 
As shown in Figure \ref{fig:evalution/number of GPUs}, the iteration time decreases with the number of GPU increases when we train a 4.3B model with a fixed global batch size.
% The speedup of \system decreases as the number of GPUs increases, where 
Specifically, \system takes 5.41s to complete an iteration on average using a GPU and 4.12s using 8 GPUs.
As the training scale increases, inter-GPU communication becomes the new bottleneck that limits the performance of \system.
However, \system still ensures a better performance when training with multi-GPUs.
For example, when scaling to training with 8 GPUs, \system increases the training speed by 28.1\% compared to DeepSpeed and 19.1\% compared to Native-Offload.

\textbf{Scalability to batch size.}
As training can be conducted with different batch sizes for best performance, we study \system by assigning larger batch sizes to a 6.7B model training in 8 GPUs.
As shown in Figure \ref{fig:evalution/batch size GPU}, we observe that \system always outperforms all the baselines when the batch size increases.
Specifically, when the batch size is 128, \system and \systemMem trains faster than DeepSpeed by $1.21\times$ and $1.16\times$ respectively.
This observation means that if resources are limited to training models and offloading is used, \system could always improve the training efficiency.

\subsection{Scaling to scale-up domain.}
\label{sec: eva/ Scaling to scale-up domain}

\begin{figure}
\centering
% \hspace{-0.4cm}
  \subfigure[CPU-to-NPU transfer bandwidth.]{
  \label{fig:multipath}
  \begin{minipage}{.21\textwidth}
  \centering
  \hspace{-0.3cm}
    \includegraphics[scale=0.24]{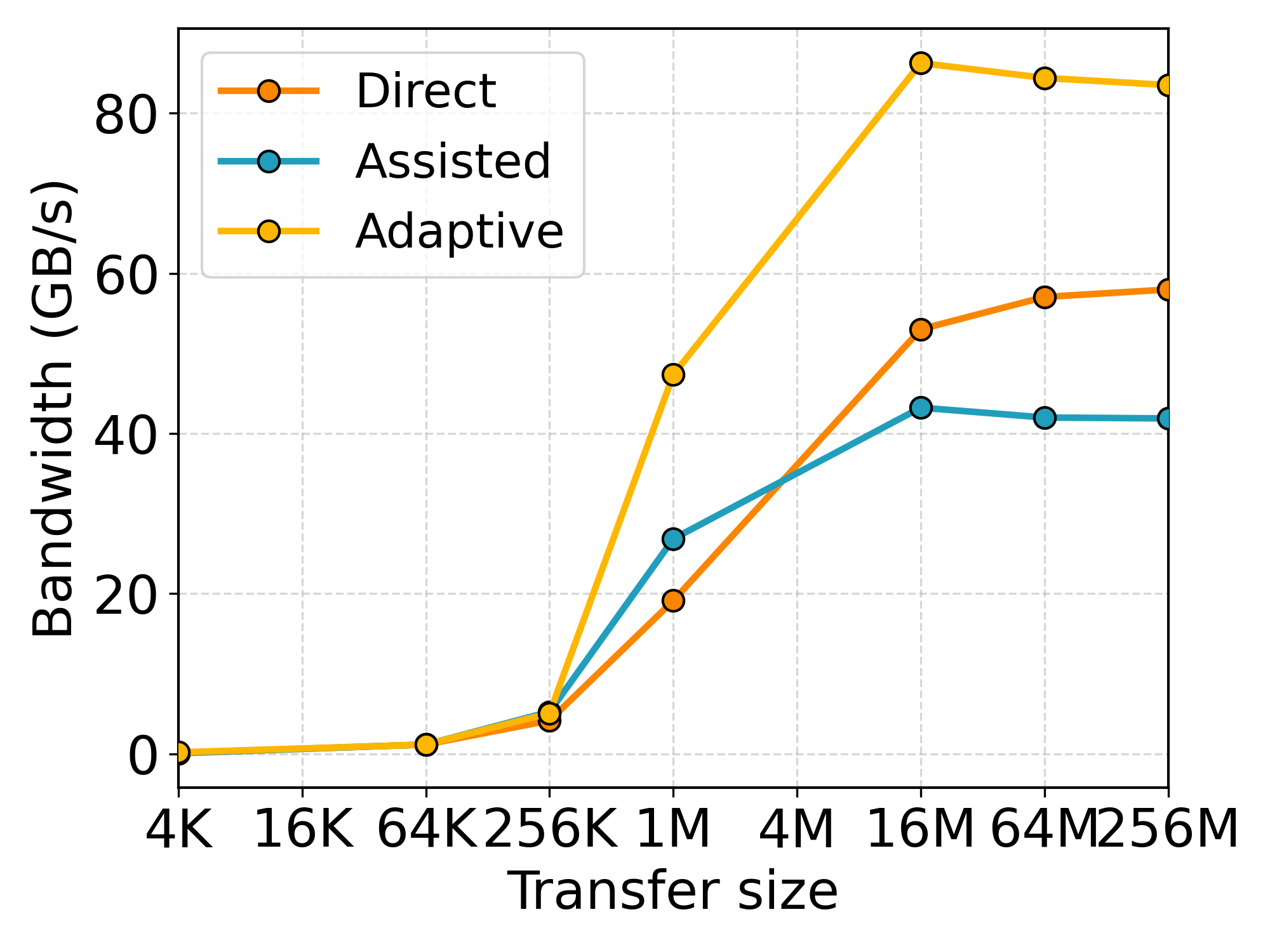}
  \end{minipage}
  }\hspace{0.4cm}
  \subfigure[End-to-end training performance.]{
  \label{fig:scaleup_e2e}
  \begin{minipage}{.21\textwidth}
  \centering
    \includegraphics[scale=0.24]{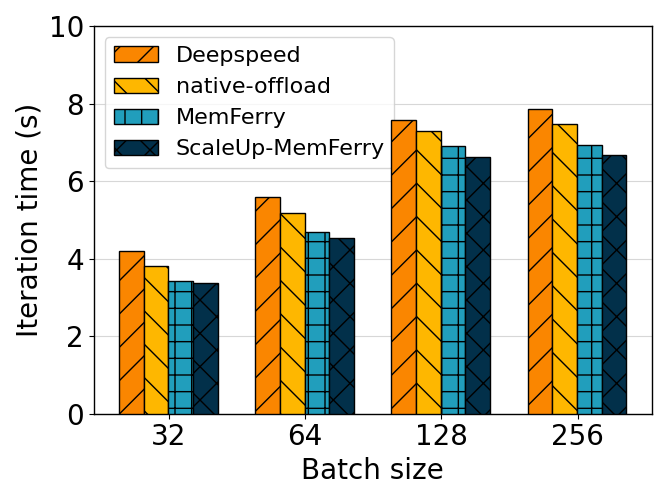}
  \end{minipage}
  }
  \vspace{-0.3cm}
  \caption{Performance of ScaleUp-MemFerry on the CloudMatrix384 scale-up architecture.}
  \label{fig:scaleup_evaluation}
  \vspace{-0.6cm}
\end{figure}

We evaluate ScaleUp-MemFerry on one compute node of Huawei CloudMatrix384, using four NPUs for model training and the other four NPUs as assisting devices for scale-up-assisted data transfer. All NPUs and CPUs are interconnected through the high-bandwidth Unified Bus (UB). We conduct two sets of experiments. First, we use a microbenchmark with one target NPU and one assisting NPU to study the CPU-to-NPU transfer performance under different transfer sizes. This experiment evaluates both the benefit of the additional assisted path and the effectiveness of our adaptive data dispatcher. Second, we evaluate the end-to-end training performance using a 13B Transformer-XL model on four NPUs. We compare DeepSpeed, Native-Offload, the original MemFerry, and ScaleUp-MemFerry, where ScaleUp-MemFerry augments MemFerry with the scale-up-assisted data transfer and adaptive data dispatcher described in Sec.~\ref{sec:design/scale-up}.

% We next evaluate the flexible gradient offload strategy of \systemMem. 
\textbf{Effectiveness of adaptive data transfer.} Figure~\ref{fig:multipath} shows the achieved CPU-to-NPU bandwidth under different transfer sizes. \emph{Direct} transfers all data through the original CPU$\rightarrow$NPU$_{\mathrm{target}}$ path, while \emph{Assisted} transfers all data through the additional CPU$\rightarrow$NPU$_{\mathrm{assist}}\rightarrow$NPU$_{\mathrm{target}}$ path. \emph{Adaptive} uses the two paths concurrently and partitions the data between them according to our adaptive data dispatcher. For small transfers, the bandwidth difference among the three schemes is limited because the transfer time is largely dominated by communication latency. As the transfer size increases, Adaptive can effectively exploit the additional bandwidth provided by the assisted path and considerably outperform either path used alone. For example, with a 16 MB transfer, Direct and Assisted achieve approximately 53 GB/s and 43 GB/s, respectively, whereas Adaptive reaches approximately 86 GB/s, corresponding to $1.62\times$ and $2.0\times$ their bandwidths. Adaptive continues to sustain more than 80 GB/s for transfer sizes from 16 MB to 256 MB. These results demonstrate that neither path alone can fully utilize the available scale-up bandwidth, while our adaptive data dispatcher effectively divides data across the two paths according to their distinct latency and bandwidth characteristics.

\textbf{End-to-end training performance.} We further investigate whether the higher CPU-to-NPU transfer bandwidth translates into end-to-end training speedup. Figure~\ref{fig:scaleup_e2e} reports the per-iteration time for training the 13B Transformer-XL model with batch sizes ranging from 32 to 256. ScaleUp-MemFerry consistently achieves the shortest iteration time across all evaluated batch sizes. Compared with the original MemFerry, ScaleUp-MemFerry further reduces the iteration time by up to approximately 4.0\%, showing that scale-up-assisted data transfer provides additional performance gains even after MemFerry has overlapped a large fraction of data movement with GPU computation. Compared with DeepSpeed, ScaleUp-MemFerry reduces the iteration time by up to approximately 20.7\%. Moreover, its performance advantage is consistently maintained as the batch size increases, indicating that the additional transfer path can be effectively incorporated into the original MemFerry execution schedule. Overall, these results demonstrate that ScaleUp-MemFerry can exploit the richer communication resources in emerging scale-up domains to further accelerate offload training.

% \vspace{-0.3cm}
\section{Related work}
\label{se:related work}
\vspace{-0.1cm}
% In this section, we briefly discuss the related work of this paper along the following two categories:

\textbf{Offload Training (or heterogeneous training).} To train large-scale models with limited GPU memory \cite{DBLP:conf/ppopp/offloadtraining1, DBLP:conf/asplos/offloadtraining2, DBLP:conf/hpca/offloadtraining3, DBLP:conf/asplos/SwapAdvisor, DBLP:conf/hpca/Sentinel, DBLP:conf/ppopp/Superneurons, DBLP:journals/tpds/iMLBench, DBLP:conf/icpp/CoTrain},
% 1) Distributed training, such as popular methods like data parallelism \cite{DBLP:conf/nips/DP,DBLP:conf/sc/zero} and model parallelism \cite{DBLP:journals/corr/megatron-lm, DBLP:conf/nips/GPipe, DBLP:conf/sosp/Pipedream, DBLP:conf/eurosys/graphPartition}. 2)  
previous works offloaded the model training content to devices with larger memory (such as CPU and NVMe \cite{DBLP:conf/fast/FlashNeuron, DBLP:conf/sc/zero-infinity, DBLP:conf/sc/STRONGHOLD}). L2L \cite{DBLP:journals/corr/L2L} offloaded parameters into CPU retrieving them as needed for computational processes during training. \system continues the approach of CPU offload while introducing GPU DHA, into training, providing a new execution schedule and gradient offloading method.

\textbf{DL Systems with Direct-Access.} In model inference, DeepPlan \cite{DBLP:conf/eurosys/DeepPlan} leveraged DHA to serve better inference, which achieves layer-wise communication overlap with computation. In deep learning training, systems \cite{DBLP:conf/glvlsi/CNN-DMA, DBLP:journals/pvldb/GNN-DMA} leveraged DMA \cite{cuda}, a GPU feature for unified cross-device data management, to reduce data copying and achieve computation-communication overlap for specific models (e.g. CNNs, GNNs). \system provides a training paradigm for general offload training by leveraging DHA not only to overlap transmission but also to reduce the training memory requirement. 

\textbf{System optimizations in scale-up networks.}
Recent studies exploit the high-bandwidth accelerator interconnects in scale-up networks to improve deep learning systems~\cite{DBLP:journals/corr/abs-2504-06095,DBLP:conf/asplos/KumarAS25,DBLP:conf/sigcomm/ShouLNM0JLXLCY025}. 
For example, Aqua~\cite{DBLP:conf/asplos/KumarAS25} uses peer GPU memory within a scale-up domain to balance inference memory usage, while InfiniteHBD~\cite{DBLP:conf/sigcomm/ShouLNM0JLXLCY025} coordinates scale-up and scale-out communication for distributed training. 
Different from these works, ScaleUp-MemFerry targets CPU-offloaded training and exploits otherwise underutilized accelerator interconnect bandwidth to assist CPU-to-accelerator data movement with adaptive multi-path transmission.

  \vspace{-0.1cm}
\section{Conclusion}
\label{sec:conclusion}
\vspace{-0.1cm}
In this paper, we propose \system, which leverages hybrid GPU execution approaches to accelerate offload training and reduce memory usage. Our key contribution is to prove that GPU's DHA feature can potentially benefit deep learning training. We integrate the GPU DHA feature into offload training and design hybrid training patterns that can fully overlap the transmission of offload training. We propose a unified management abstraction called the shadow model which removes redundant memory copying. We further propose a gradient offload strategy - \systemMem which reduces training memory requirement on demand. Evaluations show that \system can speed up offload training meanwhile significantly reducing the memory requirement for training.

% \footnotesize IEEEtranS
% \bibliographystyle{IEEEtran}
% \bibliography{IEEEabrv, sample}

% \input{body/appendix}

{ \bibliographystyle{IEEEtran}
\bibliography{IEEEabrv, sample}}

\end{document}